\documentclass[preprint, secnumarabic, amssymb, amsmath, aps, prd, nofootinbib]{revtex4-1}

\newcommand{\bra}[1]{\left\langle #1 \right|}
\newcommand{\ket}[1]{\left| #1 \right\rangle}

\newcommand{\ul}[1]{\underline{#1}}
\newcommand{\psibar}{\overline{\psi}}
\newcommand{\ubar}[1]{\overline{U}_{#1}}
\newcommand{\Tr}{\mathrm{Tr}}
\newcommand{\braket}[2]{\left\langle #1 \right| \left. #2 \right\rangle}
\newcommand{\ketbra}[2]{\ket{#1} \!\! \bra{#2}}

\newcommand{\half}{\frac{1}{2}}
\newcommand{\thalf}{\tfrac{1}{2}}
\newcommand{\ord}[1]{\mathcal{O}\left( #1 \right)}
\newcommand{\rfeq}{\overset{R.F.}{=}}
\newcommand{\crfeq}{\overset{C.R.F.}{=}}
\newcommand{\gpm}{g_{+-}}
\newcommand{\gPM}{g^{+-}}

\usepackage{graphics, graphicx,slashed}
\usepackage{color}
\usepackage{hyperref} 

\newcommand{\as}{\alpha_s}

\def\eq#1{{Eq.~(\ref{#1})}}
\def\fig#1{{Fig.~\ref{#1}}}

\begin{document}

\title{Calculating TMDs of a Large Nucleus: \\ Quasi-Classical
  Approximation and Quantum Evolution}

%\title{Calculating TMDs of an Unpolarized Target: \\ Quasi-Classical
%  Approximation and Quantum Evolution}
%\title{TMD Mixing Due to Spin-Orbit Correlations in a Heavy Nucleus}%
%
\author{Yuri V. Kovchegov}
	\email[Email: ]{kovchegov.1@osu.edu}
	\affiliation{Department of Physics, The Ohio State University, Columbus, OH 43210, USA}
\author{Matthew D. Sievert}
	\email[Email: ]{msievert@bnl.gov}
	\affiliation{Bldg. 510A, Physics Department, Brookhaven National Laboratory, Upton, NY 11973, USA}
\begin{abstract}
  We set up a formalism for calculating transverse-momentum-dependent
  parton distribution functions (TMDs) of a large nucleus using the
  tools of saturation physics.  By generalizing the quasi-classical
  Glauber-Gribov-Mueller/McLerran-Venugopalan approximation to allow
  for the possibility of spin-orbit coupling, we show how any TMD can
  be calculated in the saturation framework.  This can also be applied
  to the TMDs of a proton by modeling it as a large ``nucleus.''  To
  illustrate our technique, we calculate the quark TMDs of an
  unpolarized nucleus at large-$x$: the unpolarized quark distribution
  and the quark Boer-Mulders distribution. We observe that spin-orbit
  coupling leads to mixing between different TMDs of the nucleus and
  of the nucleons. We then consider the evolution of TMDs: at
  large-$x$, in the double-logarithmic approximation, we obtain the
  Sudakov form factor. At small-$x$ the evolution of
  unpolarized-target quark TMDs is governed by BK/JIMWLK evolution,
  while the small-$x$ evolution of polarized-target quark TMDs appears
  to be dominated by the QCD Reggeon.
\end{abstract}
\date{\today}

\maketitle

%######################################################################################
%######################################################################################

%***********************************************************************************
%\begin{figure}[hbt]
 %\centering
 %\includegraphics[width=\textwidth]{Figs/TITLE.eps}
 %\caption{}
%\label{}
%\end{figure}
%***********************************************************************************

%######################################################################################
%######################################################################################

%|||||||||||||||||||||||||||||||||||||||||||||||||||||||||||||||||||||||||||||||||||

\section{Introduction}
\label{sec-Intro}

%|||||||||||||||||||||||||||||||||||||||||||||||||||||||||||||||||||||||||||||||||||

Over the past decade quark and gluon transverse momentum-dependent
parton distribution functions (TMDs)
\cite{Collins:1989gx,Collins:1981uk} have become an integral component
of our understanding of the momentum-space structure of the nucleon.
At the same time the main principles for calculating TMDs in the
perturbative QCD framework have remained essentially the same over the
years: one parameterizes the initial conditions at some initial
virtuality $Q^2 = Q_0^2$ and then applies the Collins-Soper-Sterman
(CSS) evolution equation \cite{Collins:1989gx} to find the TMDs at all
$Q^2$. The initial conditions are non-perturbative, and have to be
constructed using models of the non-perturbative QCD dynamics (see
\cite{Bacchetta:2011bn} and references therein). Since very little is
known about non-perturbative effects in QCD, often one uses the same
form for the parameterization of the initial conditions for several
different TMDs, frequently assuming a Gaussian dependence of the TMDs
on the parton transverse momentum $k_T$ \cite{Anselmino:2007fs}. It is
clearly desirable to have a better control of our qualitative and
quantitative understanding of TMDs.

To this end one can employ the recent progress in our understanding of
small-$x$ physics and parton saturation
\cite{Gribov:1984tu,Iancu:2003xm,Weigert:2005us,Jalilian-Marian:2005jf,Gelis:2010nm,Albacete:2014fwa,KovchegovLevin}
to put constraints on the TMDs and even calculate them in the
high-energy limit. When calculating TMDs one often works either in the
$s \sim Q^2 \gg k_T^2$ (large-$x$) or in the $s \gg Q^2 \gg k_T^2$
(small-$x$) regimes, where $s$ is the center-of-mass energy and $x =
Q^2/(s+Q^2)$ if one neglects the proton mass. In either case the
energy $s$ is large, and the techniques of high-energy QCD should
apply. The degrees of freedom in saturation physics are infinite
Wilson lines along (almost) light-like paths. The definition of the
TMDs involves light-cone Wilson lines as well
\cite{Boer:2011xd,Boer:2002ju}, although the integration paths are
semi-infinite, forming the so-called ``light-cone staple''. We can see
that there are both similarities and differences between saturation
physics and the physics of TMDs. The interface of these two sub-fields
of quantum chromodynamics (QCD) has been explored in
\cite{Boer:2006rj,Boer:2008ze,Boer:2002ij,Dominguez:2011br,Metz:2011wb,Kovchegov:2012ga,Kang:2011ni,Kang:2012vm,Schafer:2013mza,Zhou:2013gsa,Altinoluk:2014oxa}.

In the past some success has been achieved in applying saturation
physics to study the Sivers function \cite{Sivers:1989cc,
  Sivers:1990fh} both in semi-inclusive deep inelastic scattering
(SIDIS) and in the Drell-Yan process (DY). In \cite{Kovchegov:2013cva}
the Sivers function was constructed by generalizing the
quasi-classical Glauber--Gribov--Mueller (GGM) \cite{Mueller:1989st}
/McLerran--Venugopalan (MV)
\cite{McLerran:1993ka,McLerran:1994vd,McLerran:1993ni} approximation
of a heavy nucleus with atomic number $A \gg 1$. The presence of the
atomic number generates a resummation parameter $\as^2 \, A^{1/3}$
\cite{Kovchegov:1996ty,Kovchegov:1997pc} allowing a systematic
resummation of multiple rescatterings, which are essential for the
Sivers function.  This picture can also be applied to the proton if
one models it as a large ``nucleus.'' This large-nucleus approximation
is known to work well in describing the data from deep inelastic
scattering (DIS) experiments on a proton at low-$x$; it is therefore
possible that it would give a reasonable description for proton TMDs
as well. The result of \cite{Kovchegov:2013cva} was an explicit form
of the Sivers function in the $s \sim Q^2 \gg k_T^2$ regime, which was
different from a simple Gaussian in $k_T$ and which can be used as the
initial condition for its Collins--Soper--Sterman (CSS) evolution
\cite{Collins:1989gx}. Another important result of
\cite{Kovchegov:2013cva} was the realization that the Sivers function
can be produced via two different channels: one is the standard
``lensing'' mechanism of
\cite{Brodsky:2002cx,Brodsky:2002rv,Brodsky:2013oya} with additional
momentum broadening due to multiple rescatterings in the nucleus,
while the other mechanism was due to the orbital angular momentum
(OAM) of the nucleus combined with multiple rescatterings. This latter
channel has not been reported before \cite{Kovchegov:2013cva}, and it
dominates mainly in the regime where multiple rescatterings are
important, or, more precisely, for $k_T \lesssim Q_s/\sqrt{\as}$,
where $Q_s$ is the saturation scale of the nucleus. It appears that
applications of saturation physics to the calculation of TMDs may lead to
qualitatively new channels of generating the relevant observables.

The aim of this work is twofold. First of all we want to generalize
the approach of \cite{Kovchegov:2013cva} to the calculation of any TMD
in the quasi-classical approximation and for $s \sim Q^2 \gg
k_T^2$. This is accomplished in Sec.~\ref{classics} for the case of an
unpolarized nucleus; the generalization to the polarized case is
straightforward and is left for future work.  The quasi-classical TMD
calculation is accomplished using the factorization given in
\eq{e:QCFact2} (and, in more detail in \eq{e:QCFact3}), which is
constructed by analogy with the particle production cross section: the
TMD is a convolution of the classical Wigner distribution $W$
describing the nucleons in the nucleus with the TMD of one of the
nucleons ($\phi^N$) and the Wilson lines $S$ of the ``staple''
evaluated in the GGM/MV approximation. The functional form of the
nucleon Wigner distribution in the nucleus can be constrained using
the symmetries of the nuclear ground state. Considering for simplicity
an unpolarized nucleus we arrive at the parameterization of
\eq{W_param}. With the help of this unpolarized nuclear Wigner
distribution we construct the unpolarized quark distribution in
\eq{e:f1-1} and the Boer-Mulders distribution \cite{Boer:1997nt} in
\eq{e:BM-3}. Note that the nuclear effects on the TMDs are not limited
to the $k_T$-broadening as is commonly assumed in the literature
\cite{Liang:2008vz}. Other TMDs for the (transversely or
longitudinally) polarized nucleus and for the unpolarized nucleus
could be constructed by analogy, but we limit the examples to the two
TMDs mentioned above along with the Sivers distribution constructed in
\cite{Kovchegov:2013cva}.

An interesting consequence of the calculation is a mixing between
different TMDs of the nucleus and of the nucleons. For instance, as
was already observed in \cite{Kovchegov:2013cva}, the nuclear Sivers
function is a linear superposition of the nucleon Sivers function and
the unpolarized quark distribution of the nucleon. Similarly, we find
that the unpolarized quark distribution of a nucleus in \eq{e:f1-1} is
a linear superposition of contributions due to the unpolarized
nucleon quark distribution and the Sivers function of the
nucleons. The nuclear Boer-Mulders function in \eq{e:BM-3} receives
contributions from the nucleon Boer-Mulders function, transversity and
pretzelosity. Such TMD mixing means that, in the quasi-classical
approximation, different nuclear TMDs are expressed in terms of the
same Wigner distributions and various nucleon TMDs. This may lead to
relations between different nuclear TMDs which, in principle, could be
tested experimentally.

The second aim of this work is to understand TMD evolution using the
methods of saturation physics. In Sec.~\ref{Evolution} we consider TMD
evolution at large and small values of Bjorken-$x$. The large-$x$
regime, corresponding to $s \sim Q^2 \gg k_T^2$, is considered in
Sec.~\ref{large-x}. There we show that the TMD evolution can be
thought of as the evolution of the light-cone Wilson lines in the
``staple'' (if one works in the light-cone gauge of the probe), as
shown in \fig{evolution}. In the large-$x$ regime the Wilson lines
combine to form an operator corresponding to the propagation of a
(quark or gluon) dipole from the nucleus to positive light-cone infinity
(SIDIS) or from negative infinity to the nucleus (DY). (A fundamental
semi-infinite dipole is used for quark TMDs, while a semi-infinite
adjoint gluon dipole should be used for the gluon TMDs.) In the
double-logarithmic approximation resumming powers of $\as \, \ln^2
(Q^2/k_T^2)$ we recover the standard Sudakov form factor
\cite{Sudakov:1954sw} which one also obtains from the CSS evolution.

In the low-$x$ regime ($s \gg Q^2 \gg k_T^2$) the diagrams giving the
leading contribution to the TMDs are slightly different from the
large-$x$ graphs, as illustrated e.g. in \fig{unpol_lowx}. For
unpolarized-target quark TMDs in the leading single-logarithmic
approximation, which resums powers of $\as \, \ln (1/x)$, we show in
Sec.~\ref{small-x} that TMD evolution is driven by the evolution of a
fundamental dipole scattering amplitude on a nucleus, that is by the
evolution of the dipole made out of infinite Wilson lines. The
evolution of this object is well-known: in the large-$N_c$ limit it is
given by the Balitsky--Kovchegov evolution equation
\cite{Balitsky:1996ub,Balitsky:1998ya,Kovchegov:1999yj,Kovchegov:1999ua},
while beyond that limit it can be found using the
Jalilian-Marian--Iancu--McLerran--Weigert--Leonidov--Kovner (JIMWLK)
evolution equation
\cite{Jalilian-Marian:1997dw,Jalilian-Marian:1997gr,Iancu:2001ad,Iancu:2000hn}. For
the polarized-target TMDs the situation is more subtle, as the
evolution has to carry the polarization information. We argue that
polarized quark TMDs at small-$x$ are governed by the nonlinear
version \cite{Itakura:2003jp} of the QCD Reggeon evolution equations
\cite{Kirschner:1983di,Kirschner:1994vc,Kirschner:1994rq,Griffiths:1999dj,Bartels:1995iu,Bartels:1996wc,Bartels:2003dj}
(see \fig{lowx_pol} and \eq{eq:Revol}), which employs the dipole
amplitude obtained from the BK/JIMWLK equations.

We have thus constructed quasi-classical expressions for TMDs at
large-$x$ in the GGM/MV approximation, and constructed evolution of
TMDs at both large- and small-$x$. The results of this work can be
used for constructing a global TMD fit.

%-----------------------------------------------------------------------------------

%\subsection{Physics Background}

%-----------------------------------------------------------------------------------

%-----------------------------------------------------------------------------------

%\subsection{Preliminaries}

%-----------------------------------------------------------------------------------

%######################################################################################
%######################################################################################

\section{Quasi-Classical Initial Conditions}
\label{classics}

In this Section we expand and generalize the approach developed in
\cite{Kovchegov:2013cva}. While the quasi-classical method detailed
below was used in \cite{Kovchegov:2013cva} to calculate the Sivers
function of a large nucleus, here we will apply it to calculating TMDs
of an unpolarized nucleus, such as the unpolarized quark distribution
and the Boer-Mulders function.

%|||||||||||||||||||||||||||||||||||||||||||||||||||||||||||||||||||||||||||||||||||

\subsection{Quasi-Classical Factorization of the TMD Quark Distribution of a Heavy Nucleus}
\label{sec-QCfact}

%|||||||||||||||||||||||||||||||||||||||||||||||||||||||||||||||||||||||||||||||||||

%-----------------------------------------------------------------------------------

\subsubsection{The TMD Decomposition}

%-----------------------------------------------------------------------------------

The transverse-momentum-dependent (TMD) quark correlation function in
a hadronic state $\ket{h(P, S)}$ with momentum $P$ and spin $S$ is
defined by
\begin{align} \label{e:qcorr1} \phi_{\alpha \beta} (x, \ul{k}; P, S)
  \equiv \frac{g_{+-}}{(2\pi)^3} \int d^{2-}r \, e^{i k \cdot r} \,
  \bra{h (P, S)} \, \psibar_\beta (0) \, \mathcal{U}[0,r] \,
  \psi_\alpha (r) \ket{h (P, S)}_{r^+ = 0}
\end{align}
where $\alpha , \beta$ are Dirac indices, the separation vector
between the quark fields is $r^\mu = (0^+, r^-, \ul{r})$, and the
``staple-shaped'' gauge link $\, \mathcal{U}[0,r]$ extends to
future/past light-cone infinity $(\pm \infty^-)$ depending on the
process under consideration. The abbreviated notation for the
integration measure is $d^{2-}r = d^2 r_\perp \, d r^-$.  We work in
light-front coordinates
\begin{align}
  x^\pm \equiv \sqrt{\frac{\gPM}{2}} \left( x^0 \pm x^3 \right)
\end{align}
where $\gPM = 1$ and $\gPM = 2$ are two common choices of the
light-front metric. We will use a frame in which the nucleus or the
proton have a momentum predominantly in the $x^+$-direction.

Projecting the TMD correlator \eqref{e:qcorr1} onto various Dirac
matrices gives the distribution of quarks with zero, longitudinal, or
transverse polarizations, and the independent spin-spin and spin-orbit
correlations are parameterized in terms of boost-invariant TMD parton
distribution functions (the ``TMDs'').  At leading power in the large
momentum $P^+$, the TMD decomposition can be written as (see,
e.g. \cite{Boer:1997nt, Mulders:1995dh, Meissner:2007rx})
\begin{align} \label{e:qTMD1} \phi(x,\ul{k};P,S) &= \left(f_1 -
    \frac{\ul{k} \times \ul{S}}{m} f_{1T}^\bot \right)
  \left[\frac{1}{2} g_{+-} \gamma^- \right] + \left(S_L g_1 +
    \frac{\ul{k} \cdot \ul{S}}{m} g_{1T} \right) \left[\frac{1}{2}
    g_{+-} \gamma^5 \gamma^- \right] + \\ \nonumber
  & \!\!\!\!\!\!\!\!\!\!\!\!\!\!\!\!\!\!\!\!\!\!\!\! + \left(S_\bot^i
    h_{1T} + \frac{k_\bot^i}{m} S_L h_{1L}^\bot + \frac{k_\bot^i}{m}
    \frac{(\ul{k} \cdot \ul{S})}{m} h_{1T}^\bot \right)
  \left[\frac{1}{2} g_{+-} \gamma^5 \gamma_{\bot i} \gamma^- \right] +
  \left(\frac{k_\bot^i}{m} h_1^\bot \right) \left[\frac{i}{2} g_{+-}
    \gamma_{\bot i} \gamma^- \right] ,
\end{align}
where $(\ul{S}, S_L)$ denote the transverse and longitudinal
components of the spin vector, $m$ is the hadron mass, and $\ul{k}
\times \ul{S} \equiv k_x \, S_y - k_y \, S_x$ with $(x,y)$ the
coordinates in the plane transverse to the beam. The eight
leading-twist quark TMDs shown here 
\begin{align}
  \label{eq:quark_TMDs}
  \{f_1 \,,\, f_{1T}^\bot \,,\, g_1 \,,\, g_{1T} \,,\, h_{1T} \,,\,
  h_{1L}^\bot \,,\, h_{1T}^\bot \,,\, h_1^\bot\}
\end{align}
are functions of $x$ and $k_T = |\ul{k}|$.

At lowest order in $\alpha_s$ such that the gauge link
$\mathcal{U}[0,r]$ can be neglected, the correlator \eqref{e:qcorr1}
has a simple interpretation as the expectation value of the quark
density in the hadronic state $\ket{h(P, S)}$:
\begin{align} 
  \label{e:qdens}
  \phi_{\alpha \beta} (x,\ul{k}; P, S) \overset{L.O.}{=}
  \frac{1}{4(2\pi)^3 \Omega} \, \frac{1}{x k^+} \sum_{\sigma \sigma'}
  \bra{h(P,S)} b_{k \sigma'}^\dagger b_{k \sigma} \ket{h(P, S)} \left[
    \left(\ubar{\sigma'}(k)\right)_\beta \, \left(U_\sigma (k)
    \right)_\alpha \right],
\end{align}
where $\Omega \equiv g_{+ -} p^+ \int d^{2-} b$ is a boost-invariant
volume factor which normalizes the plane-wave states $\ket{h(P,S)}$.
The gauge link $\mathcal{U}[0,r]$ is needed to make the definition
gauge invariant, and it reflects the distortion of the quark
distribution in the presence of initial- or final-state gauge fields
\cite{Burkardt:2010sy, Burkardt:2008ps}.

By tracing \eqref{e:qdens} with one of $\Gamma \in \{ \gamma^+ \,,\,
\gamma^+ \gamma^5 \,,\, \gamma^5 \gamma^+ \gamma_\bot^j \}$, we
project out the distributions corresponding to unpolarized quarks
($U$), longitudinally-polarized quarks ($L$), and
transversely-polarized quarks in direction $\hat{j}$ (labeled $T^j$),
respectively.  This can be explicitly verified by evaluating the
associated spinor products in \eqref{e:qdens} using the spinor
conventions of \cite{Lepage:1980fj}:
\begin{align} \label{e:proj1}
 U \, : \hspace {1cm}
 \ubar{\sigma'} (k) \, \gamma^+ \, U_\sigma (k) &= 2 k^+
  \begin{bmatrix}
	 1 & 0 \\ 0 & 1
       \end{bmatrix}_{\sigma' \sigma} = 2 k^+ \left[ \mathbf{1}
       \right]_{\sigma' \sigma}
 \\
 L \, : \hspace {0.7 cm}
 \ubar{\sigma'} (k) \, \gamma^+ \gamma^5 \, U_\sigma (k) &= 2 k^+
  \begin{bmatrix}
    1 & 0 \\ 0 & -1
       \end{bmatrix}_{\sigma' \sigma} = 2 k^+ \left[ \sigma^3
       \right]_{\sigma' \sigma}
 \\
 T^j \, : \hspace {0.1 cm} \ubar{\sigma'} (k) \, \gamma^5 \gamma^+
 \gamma_\bot^j \, U_\sigma (k) &= 2 k^+
  \begin{bmatrix}
	 0 & \delta^{j 1} - i \delta^{j 2} \\ \delta^{j 1} + i \delta^{j 2} & 0
	\end{bmatrix}_{\sigma' \sigma}
 \\ \nonumber 
 &= 2 k^+ \left( \delta^{j 1} \left[ \sigma^1 \right]_{\sigma' \sigma}
   + \delta^{j 2} \left[ \sigma^2 \right]_{\sigma' \sigma} \right) .
\end{align}
These projections select out the corresponding set of TMDs from the
decomposition \eqref{e:qTMD1},
\begin{align} \label{e:qTMD2} &U: \phi^{[\gamma^+]} \equiv \frac{1}{2}
  \Tr\left[\phi \, \gamma^+ \right] = f_1 -
  \frac{\ul{k}\times\ul{S}}{m} f_{1T}^\bot \\ \nonumber &L:
  \phi^{[\gamma^+ \gamma^5]} \equiv \frac{1}{2} \Tr\left[\phi \,
    \gamma^+ \gamma^5 \right] = S_L g_1 + \frac{\ul{k}\cdot\ul{S}}{m}
  g_{1T} \\ \nonumber &T^j: \phi^{[\gamma^5 \gamma^+ \gamma_\bot^j]}
  \equiv \frac{1}{2} \Tr\left[\phi \, \gamma^5 \gamma^+ \gamma_\bot^j
  \right] = S_\bot^j h_{1T} + \frac{k_\bot^j}{m} S_L h_{1L}^\bot +
  \frac{k_\bot^j}{m} \frac{(\ul{k}\cdot\ul{S})}{m} h_{1T}^\bot +
  \epsilon_T^{j i} \frac{k_\bot^i}{m} h_1^\bot ,
\end{align}
so that $f_1$ represents the azimuthally symmetric distribution of
unpolarized quarks in an unpolarized hadron, the Sivers function
$f_{1T}^\bot$ represents the azimuthally asymmetric distribution of
unpolarized quarks in a transversely polarized hadron, and so on.

%-----------------------------------------------------------------------------------

\subsubsection{Quasi-Classical Factorization in a Heavy Nucleus}

%-----------------------------------------------------------------------------------

As derived in \cite{Kovchegov:2013cva}, a heavy nucleus with $A \gg 1$
nucleons in the quasi-classical approximation $\alpha_s^2 A^{1/3} \sim
\ord{1}$ admits a decomposition of its TMD quark correlator $\Phi^A$
in terms of the TMD quark correlator $\phi^N$ of its nucleons, Wilson
lines, and the nuclear Wigner distribution.  The heavy nucleus in this
parametric limit justifies two powerful simplifications of the
definition \eqref{e:qcorr1} as follows.  First, the large number of
nucleons justifies a mean-field description of the nucleus
$\ket{A(P,S)}$ in terms of the light-front wave functions of
single-nucleon states $\ket{N(p,\sigma)}$.  Second, because of the
large number of nucleons, the initial- / final-state rescattering
described by the gauge link $\mathcal{U}[0,r]$ is more likely to occur
on the many $(A-1)$ spectator nucleons, rather than on the same
nucleon from which the quark distribution is taken. Schematically, we
can write this as
\begin{align} \label{e:QCFact0} \bra{A} \psibar_{\beta} (0) \,
  \mathcal{U}[0,r] \, \psi_\alpha (r) \ket{A} \approx (\Psi_N \,
  \Psi_N^*) \, \times \, \bra{N} \, &\psibar_{\beta} (0) \, u [0,r] \,
  \psi_\alpha (r) \ket{N}
  \notag \\ & \times \,
   \bra{A-1}  \mathcal{U}[0,r] \ket{A-1} + \ord{A^{-1/3}}.
\end{align}
The resummation $\alpha_s^2 A^{1/3} \sim \ord{1}$
\cite{Kovchegov:1997pc} is a systematic way of calculating the
multiple rescattering in the weak-coupling saturation framework, and
the accuracy to leading order in $A^{-1/3}$ means that the sensitivity
of the gauge link $u [0,r]$ to the active nucleon is limited to
$\ord{\alpha_s}$.  In this way, the TMD correlator \eqref{e:qcorr1} is
factorized into a convolution of 3 factors: a wave function piece
describing the distribution of nucleons in the nucleus, the TMD
correlator of the nucleon itself, and a piece describing the multiple
rescattering of the gauge link on the spectator nucleons.  This
essential picture should apply not only to a heavy nucleus, but to any
system - such as a proton at high energies - in which the density of
color charges becomes sufficiently large.

%***********************************************************************************
\begin{figure}[hbt]
 \centering
 \includegraphics[width=0.8\textwidth]{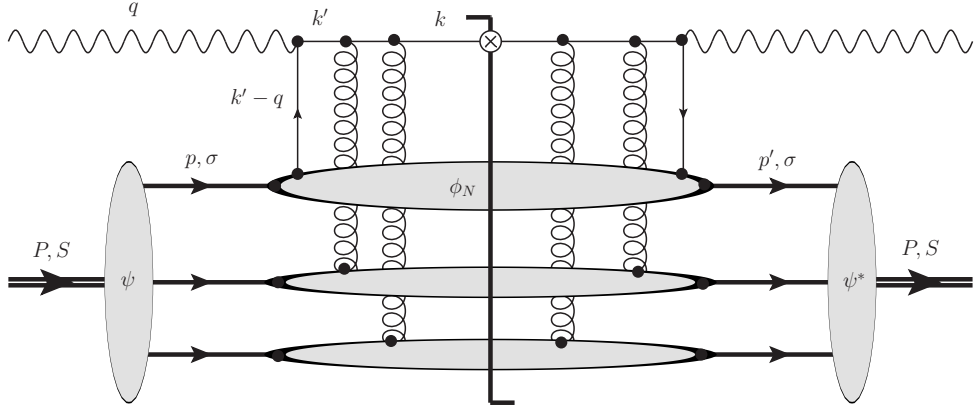}
 \caption{Quasi-classical factorization of semi-inclusive deep
   inelastic scattering on a heavy nucleus.}
\label{f:QCFact}
\end{figure}
%***********************************************************************************

For semi-inclusive deep inelastic scattering in Bjorken kinematics
(see Fig.~\ref{f:QCFact}), the quasi-classical factorization formula
is \cite{Kovchegov:2013cva}
\begin{align} 
  \label{e:QCFact1}
  \Phi_{\alpha \beta}^A (x, \ul{k}; P, S) &= A \frac{g_{+-}}{(2\pi)^5}
  \, \sum_\sigma \int d^{2+} p \, d b^- \, d^2 x \, d^2 y \, d^2 k' \, e^{-i
    (\ul{k} - \ul{k'}) \cdot (\ul{x} - \ul{y})} \, W_\sigma (p,b; P,
  S) \\ \nonumber
  & \times \phi_{\alpha \beta}^N (\hat{x}, \ul{k}' - \hat{x} \ul{p}; p
  , \sigma) \ D_{x y}[\infty^-, b^-],
\end{align}
where $\hat{x} \equiv x \frac{P^+}{p^+}$ is the quark momentum
fraction with respect to the active nucleon, $d^{2+} p = d^2 p_\perp
\, d p^+$, and $b^\mu
\equiv (0^+, b^-, \thalf(\ul{x} + \ul{y}))$ is the position of the
struck nucleon.  The Wigner distribution of nucleons
$\ket{N(p,\sigma)}$ inside the nucleus $\ket{A(P,S)}$ is
\begin{align}
\label{e:Wig1} 
W_\sigma (\overline{p},b;P,S) \equiv
  \frac{1}{2(2\pi)^3} & \int \frac{d^{2+}(p-p')}{\sqrt{p^+ p'^+}} \,
  e^{-i(p-p')\cdot b}
 \\ \nonumber &\times 
 \sum_{\mathbb{X}} \braket{A(P, S)}{N(p' ,\sigma) ; \mathbb{X}} \,
 \braket{N(p, \sigma) ; \mathbb {X}}{A(P, S)},
\end{align}
where $\overline{p} \equiv \tfrac{p+p'}{2}$ is the average momentum of
the struck nucleon in the amplitude and complex-conjugate amplitude
and $\sigma$ is its spin.  The Wigner distribution is normalized such that
\begin{align}
  g_{+-} \sum_\sigma \int \frac{d^{2} p_\perp \, d p^+ \, d^2 b_\perp
    \, d b^-}{(2 \pi)^3} W_\sigma (p,b; P, S) =1.
\end{align}

The semi-infinite dipole scattering amplitude $D_{x y}[\infty^-, b^-]$
describes the final-state rescattering on the fraction of nucleons at
depths greater than $b^-$ and is given in covariant gauge by
\begin{align} 
  \label{e:dipop}
  D_{x y}[\infty^-, b^-] \equiv \frac{1}{N_c} \left\langle \Tr \left[
      V_x [\infty^-, b^-] \, V_y^\dagger [\infty^-, b^-] \right]
  \right\rangle,
\end{align}
where 
\begin{align}
  V_x [\infty^-, b^-] \equiv \mathcal{P} \exp\left[ i g 
    \int\limits_{b^-}^{\infty^-} dz^- \gpm A^{+ a} (0^+, z^-, \ul{x}) \,
    T^a \right]
\end{align}
is a Wilson line in the fundamental representation and $T^a$ is the
fundamental generator of SU($N_c$). The angle brackets in \eq{e:dipop}
denote averaging in the nuclear wave function.

%%%%%%%%%%%%%%%%%%%%%%%%%%%%%%%%%%%%%%%%%%%%%%%%%%%%%%%%%%%%%%%%%%
\begin{figure}[t]
\centering
\includegraphics[width= \textwidth]{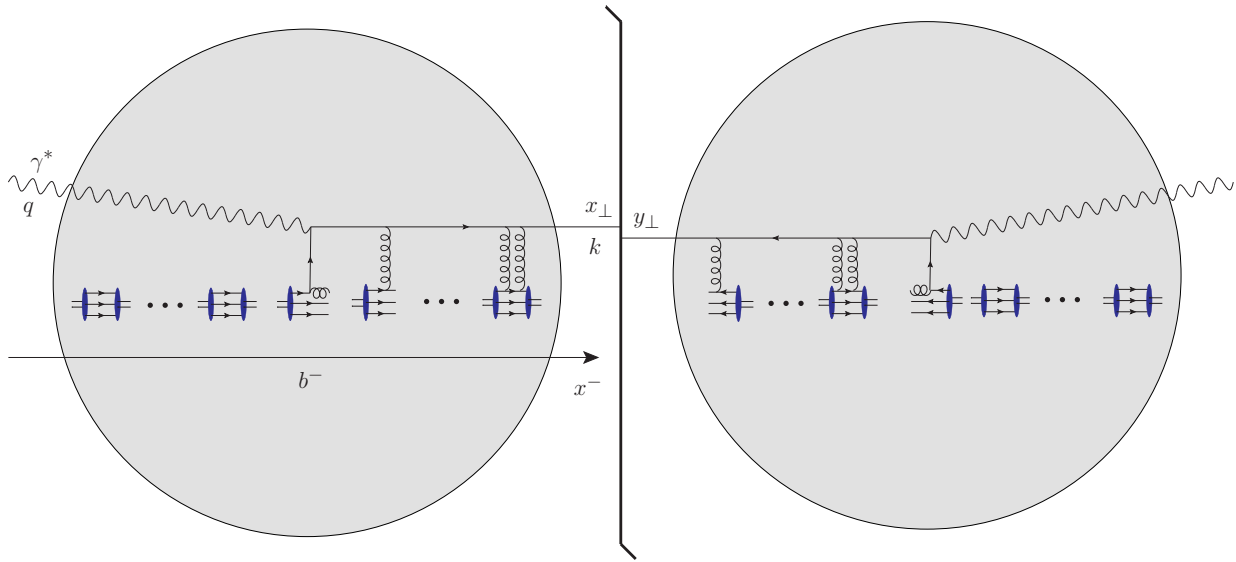}
\caption{SIDIS cross section as a square of the scattering amplitude
  explicitly illustrating the $x^-$-ordering of the nucleons in the
  nucleus. The solid vertical line denotes the final-state cut.}
\label{SIDIS_qprod}
\end{figure}
%%%%%%%%%%%%%%%%%%%%%%%%%%%%%%%%%%%%%%%%%%%%%%%%%%%%%%%%%%%%%%%%%%

The origin of the nuclear TMD decomposition \eqref{e:QCFact1} is in a
similar decomposition for the quark production cross section in the
SIDIS case \cite{Kovchegov:2013cva}, illustrated in \fig{SIDIS_qprod}
(with a similar cross-section analogy also valid in the DY
case). There the virtual photon traverses the nucleus until
interacting with one of the nucleons (the same nucleon in the
amplitude and the complex conjugate amplitude). The quark produced in
the interaction propagates through the rest of the nucleus,
interacting with the nucleons. If one works in the covariant gauge or
the $A^- =0$ light-cone gauge (with the nucleus moving along the $x^+$
light cone) the interactions of the quark with the nucleons in the
nucleus are instantaneous Coulomb gluon exchanges
\cite{Mueller:1989st} which contribute to the Wilson line describing
the quark propagator. The quark production cross section is
\cite{Kovchegov:2013cva}
\begin{align}
  \label{xsect_W3}
  \frac{d \sigma^{\gamma^* + A \to q + X}}{d^2 k \, dy} = & A \,
  g_{+-} \sum_\sigma \int \frac{d p^+ \, d^2 p \, d b^- }{(2 \pi)^3}
  \, \int d^2x \, d^2 y \ W_\sigma ( p , b; P, S ) \notag \\
  & \times \int \frac{d^2 k'}{(2 \pi)^2} \, e^{- i \,({\ul k} - {\ul
      k}') \cdot ({\ul x} - {\ul y})} \, \frac{d
    \hat{\sigma}^{\gamma^* + N \to q + X}}{d^2 k' \, dy} (p,q) \,
  D_{{x} \, {y}} [\infty^-, b^-].
\end{align}
Note that the light-cone Wilson lines describing the produced quark in
the amplitude and in the complex conjugate amplitude (in the
above-mentioned gauges) have to be at different transverse positions
$x_\perp$ and $y_\perp$ since the transverse momentum of the quark
$k_\perp$ is fixed (see e.g. \cite{Kovchegov:1998bi}). Together these
Wilson lines contribute to the ``staple'' $\mathcal{U}$ in the TMD
definition \eqref{e:qcorr1}. (The transverse link at infinity is zero
in the covariant gauge and in the $A^- =0$ light-cone gauge.)

Eq.~\eqref{e:QCFact1} can be simplified further by replacing the
dipole scattering amplitude $D_{x y}$ with its symmetric part $S_{x y}
\equiv \tfrac{1}{2} (D_{x y} + D_{y x})$ since the anti-symmetric
(odderon) part is suppressed by $A^{-1/3}$ \cite{Kovchegov:2012ga}.
If an analytic expression is desired, this symmetric part of the
dipole scattering amplitude can be evaluated in the quasi-classical
GGM/MV multiple scattering approximation to give
\cite{Mueller:1989st}
\begin{align} 
  \label{e:GGM}
  S_{x y} [\infty^-, b^-] & \equiv \frac{1}{2} \left( D_{x y}
    [\infty^-, b^-] + D_{y x} [\infty^-, b^-] \right)
 \\ \nonumber
 &= \exp\left[-\frac{1}{4} |x-y|_T^2 \, Q_s^2
   \left(\left|\tfrac{x+y}{2}\right|_T \right) \left(\frac{R^-
       (|\tfrac{x+y}{2}|_T) - b^-}{2 R^- (|\tfrac{x+y}{2}|_T)} \right)
   \ln\frac{1}{|x-y|_T \Lambda} \right] .
\end{align}
Here $Q_s (b_T)$ is the saturation scale at impact parameter $b_T$,
$R^- (b_T)$ is the longitudinal radius of the nucleus at impact
parameter $b_T$ and $(\frac{R^- (b_T) - b^-}{2 R^- (b_T)})$ is the
fraction of nucleons which participate in the final-state
rescattering.  The logarithm with infrared (IR) regulator $\Lambda$,
which is often neglected, is only important for recovering the
perturbative large-$k_T$ (small-$|x-y|_T$) asymptotics.
Alternatively, the Wilson line operators can be evaluated on a
configuration-by-configuration basis if desired using Monte Carlo
methods along the lines of \cite{Dumitru:2011vk, Rummukainen:2003ns}.
By replacing $D_{x y}$ with $S_{x y}$, we can rewrite the
quasi-classical factorization formula \eqref{e:QCFact1} as
\begin{align} 
  \label{e:QCFact2}
  \Phi_{\alpha \beta}^A (x, \ul{k}; P, S) &= A \frac{g_{+-}}{(2\pi)^5}
  \, \sum_\sigma \, \int d^{2+} p \, d^{2-} b \, d^2 r \, d^2 k' \,
  e^{-i (\ul{k} - \ul{k'} - \hat{x} \ul{p}) \cdot \ul{r}} \, \\
  \nonumber
  & \times W_\sigma (p,b; P, S) \, \phi_{\alpha \beta}^N (\hat{x},
  \ul{k}' ; p , \sigma) \, S_{(r_T, b_T)}^{[\infty^-, b^-]},
\end{align}
where we have changed variables to $\ul{b} \equiv \tfrac{\ul{x}+
  \ul{y}}{2}$ and $\ul{r} \equiv \ul{x} - \ul{y}$ and shifted the
integration variable $\ul{k^\prime} \rightarrow \ul{k^\prime} +
\hat{x} \ul{p}$.

%-----------------------------------------------------------------------------------

\subsubsection{Lorentz-Covariant Spin Decompositions}

%-----------------------------------------------------------------------------------

The quasi-classical factorization formula \eqref{e:QCFact2} contains a
sum over the spins $\sigma$ of the intermediate nucleons.  In
\cite{Kovchegov:2013cva}, a particular spin basis was chosen (the
transverse $\ket{\pm x}$ basis) and was used to evaluate this spin
sum; this effectively limited the polarizations of the intermediate
nucleons to be either unpolarized, or polarized in the direction of
the chosen basis.  As we will now show, in a more complete treatment,
the polarization of the intermediate nucleons can be arbitrary -
unpolarized, longitudinally polarized, or polarized in either of the
transverse directions - independent of the choice of spin basis.

Let us first write a decomposition of the Wigner distribution
\eqref{e:Wig1} for nucleons with an arbitrary spin state $\ket{S}$ in
terms of the light-cone helicity basis $\ket{\pm}$.  For simplicity,
we will restrict ourselves to an unpolarized nucleus, but which may
have polarized nucleons, which we write in the compact form
\begin{align} \label{e:Wig2} W (\overline{p},b, S) \equiv
  \frac{1}{2(2\pi)^3} & \int \frac{d^{2+}(\delta p)}{\sqrt{p^+
      p^{\prime +}}} \, e^{-i \delta p \cdot b} \braket{A(P)}{N(p'
    ,S)} \, \braket{N(p, S)}{A(P)},
\end{align}
where $\delta p \equiv p - p^\prime$, $\bar{p} \equiv \half( p +
p^\prime)$, and $b^\mu \equiv (0^+, b^-, \ul{b})$.  As described in
\cite{Diehl:2005jf}, an arbitrary spin state can be decomposed in
terms of the light-cone helicity basis as
\begin{align} \label{e:spin1} \ket{S} \equiv \cos\frac{\theta}{2}
  \ket{+} + \sin\frac{\theta}{2} e^{i \phi} \ket{-};
\end{align}
in the rest frame (R.F.) of the nucleon, the eigen-axis of spin
projections for this state is given by the three-dimensional spin
vector
\begin{align} \label{e:spinRF1} \vec{S} \rfeq (\sin\theta \cos\phi,
  \sin\theta \sin\phi , \cos\theta) \equiv (S_\bot^1 , S_\bot^2,
  \lambda)
\end{align}
with $\lambda^2 + S_T^2 = 1$.  The Lorentz-covariant spin vector
$S^\mu$ is obtained by defining \eqref{e:spinRF1} as the spatial part
in the rest frame and boosting it to a frame in which the nucleon
momentum is $\vec{p}$ \footnote{\eq{e:spinRF2} constitutes the
  canonical definition of spin \cite{Polyzou:2012ut,Leader:2013jra},
  obtained by boosting $(0, \vec{S}_{R.F.})^\mu$ out of the rest frame
  with a single rotationless boost.  This can also be regarded as
  defining $S^\mu \equiv W^\mu /m$ with $W^\mu$ the Pauli-Lubanski
  four-vector, and obtaining $\vec{S}_{R.F.}$ via \eq{e:spinRF2}.
  Alternatively, one may use light-front boosts, obtaining an
  expression for $S^\mu$ in terms of $\vec{S}'_{R.F.}$, which is the
  spin in a rest frame rotated with respect to the rest frame used in
  the second line of \eqref{e:spinRF2} (see \cite{Polyzou:2012ut} and
  the footnote 35 on page 213 of \cite{Leader:2013jra}).  Ultimately,
  for the situations we consider (nonrelativistic nucleon motion and
  an ultrarelativistic boost along the $+z$-axis), both definitions of
  the spin three-vector coincide.  We thank C\'{e}dric Lorc\'{e} for
  helping us clarify these subtleties.}:
\begin{align} 
  \label{e:spinRF2}
S^\mu (p) & \equiv \left(
    \frac{\vec{S}_{R.F.}\cdot\vec{p}}{m} , \vec{S}_{R.F.} +
    \frac{\vec{p}}{m} \left[ \frac{\vec{p}\cdot{\vec{S}_{R.F.}}}{E +
        m} \right] \right)
  \notag \\ & \overset{R.F.}{=}
  \left( 0, \vec{S}_{R.F.} \right) = \left( 0, S_\bot^1, S_\bot^2,
    \lambda \right).
\end{align}
Note that, by construction, $p_\mu S^\mu(p) = 0$. We have also made
use of the fact that the nucleons do not interact with each other in
the quasi-classical approximation at hand
\cite{Lorce:2011dv,Lorce:2011zta,Leader:2013jra}.

By forming the outer product of \eqref{e:spin1} and using
\eqref{e:spinRF2} , we obtain
\begin{align}
  \ketbra{S}{S} &\rfeq \thalf \bigg[ \ketbra{+}{+} \: + \:
  \ketbra{-}{-} \bigg] + \thalf \lambda \bigg[ \ketbra{+}{+} \: - \:
  \ketbra{-}{-} \bigg]
  \notag \\ & +
  \thalf S_\bot^1 \bigg[ \ketbra{-}{+} \: + \: \ketbra{+}{-} \bigg] +
  \thalf S_\bot^2 \bigg[ i \ketbra{-}{+} \: - \: i \ketbra{+}{-}
  \bigg]
  \notag \\ \ketbra{S}{S} &\rfeq
  \half
    \begin{bmatrix}
	 \ket{+} & \ket{-} 
    \end{bmatrix}
  \bigg\{ 
   \begin{bmatrix}
      \: \mathbf{1} \:
    \end{bmatrix}
  + \vec{S}_{R.F.} \cdot 
    \begin{bmatrix}
      \: \vec{\sigma} \:
    \end{bmatrix}
  \bigg\}
    \begin{bmatrix}
	 \bra{+} \\ \bra{-} 
    \end{bmatrix} ,
\end{align}
\newpage
\noindent where we have introduced the Pauli matrices and unit matrix.  If we
define a Lorentz-covariant set of Pauli matrices analogous to
\eqref{e:spinRF2}
\begin{align} 
\label{e:spinRF3} 
\hat\sigma^\mu (p) & \equiv \left( \frac{\vec{\sigma}\cdot\vec{p}}{m}
  , \vec{\sigma} + \frac{\vec{p}}{m} \left[
    \frac{\vec{p}\cdot{\vec{\sigma}}}{E + m} \right] \right)
  \notag \\ & \overset{R.F.}{=}
  \left( 0, \vec{\sigma} \right) = \left( 0, \sigma_\bot^1,
    \sigma_\bot^2, \sigma_\bot^3 \right),
\end{align}
where again $p_\mu \hat\sigma^\mu (p) = 0$ by construction, then we
have $S_\mu (p) \hat\sigma^\mu (p) = - \vec{S}_{R.F.} \cdot
\vec{\sigma}$ in any frame, and
\begin{align}
  \ketbra{S}{S} &= \half
    \begin{bmatrix}
	 \ket{+} & \ket{-} 
    \end{bmatrix}
  \bigg\{ 
   \begin{bmatrix}
      \: \mathbf{1} \:
    \end{bmatrix}
  - S_\mu (p)
    \begin{bmatrix}
      \: \hat\sigma^\mu (p) \:
    \end{bmatrix}
  \bigg\}
    \begin{bmatrix}
	 \bra{+} \\ \bra{-} 
    \end{bmatrix} .
\end{align}
Inserting this into \eqref{e:Wig2} we obtain
\begin{align} 
  \label{e:Wig3}
  W(\bar p, b, S) &= W_{unp}(\bar p,b) - S_\mu (\bar p) \,
  \hat{W}_{pol}^\mu (\bar p , b)
\end{align}
with
\begin{align} 
  \label{e:Wig4}
  W_{unp} (\overline{p},b) &= \half \sum_{\lambda \lambda'} \bigg[
  W_{\lambda \lambda'}(\bar p,b) \: \mathbf{1}_{\lambda' \lambda}
  \bigg] = \half \Tr[ W(\bar p,b) ]
 \notag \\
 \hat{W}_{pol}^\mu (\bar p , b) &=
 \half \sum_{\lambda \lambda'} \bigg[ W_{\lambda \lambda'}(\bar p,b)
 \: \hat\sigma^\mu_{\lambda' \lambda} (\bar p) \bigg] = \half \Tr[
 W(\bar p,b) \: \hat\sigma^\mu (\bar p)],
\end{align}
where we have also introduced the Hermitean $( 2\times 2)$ matrix
\begin{align} 
  \label{e:Wigmat}
  W_{\lambda \lambda'} (\overline{p},b) &\equiv \frac{1}{2(2\pi)^3}
  \int \frac{d^{2+}(\delta p)}{\sqrt{p^+ p^{\prime +}}} \, e^{-i
    \delta p \cdot b} \braket{A(P)}{N(p', \lambda)} \, \braket{N(p,
    \lambda')}{A(P)}.
\end{align}

The projections $W_{unp}$ and $- S_\mu \hat{W}_{pol}^\mu$ select out
the Wigner distributions of nucleons which have zero polarization, or
polarization $S^\mu$, respectively.  And although they were derived
here in terms of the light-cone helicity basis $\ket{\pm}$, they are
invariant under a change of basis; any unitary rotation of the spin
states transforms the matrix $W_{\lambda \lambda'}$, but it also
rotates the Pauli matrices such that the traces \eqref{e:Wig4} remain
invariant.
We can also write a decomposition of the matrix $W_{\lambda \lambda'}$
in terms of the complete basis $\{\mathbf{1} , \vec{\sigma} \}$ to
invert Eqs.~\eqref{e:Wig3}, \eqref{e:Wig4}:
\begin{align} 
\label{e:Wig5} 
W_{\lambda \lambda'} (\bar p, b) &=
  W_{unp} (\bar p, b) \: \bigg[ \:\: \mathbf{1} \:\: \bigg]_{\lambda
    \lambda'} - \hat{W}_{pol, \, \mu} (\bar p, b) \: \bigg[ \:
  \hat\sigma^\mu (\bar p) \: \bigg]_{\lambda \lambda'}.
\end{align}
From \eqref{e:spinRF3} we obtain the trace identity
\begin{align} \label{e:spinRF4} -\half\Tr[\hat\sigma^\mu (p) \,
  \hat\sigma^\nu (p) ] = g^{\mu \nu} - \frac{p^\mu p^\nu}{m^2}
\end{align}
which allows us to recover \eqref{e:Wig4} from \eqref{e:Wig5} by
tracing the latter with $\hat\sigma^\nu$ or $\mathbf{1}$ and using the
fact that $\bar{p}_\mu \hat{W}_{pol}^\mu = \half \Tr[ W \: \bar{p}_\mu
\hat\sigma^\mu (\bar p)] = 0$.

Note that the normalization of the matrix Wigner distribution
$W_{\lambda \lambda'} (p, b)$ is
\begin{align}
  g_{+-} \int \frac{d^{2} p_\perp \, d p^+ \, d^2 b_\perp \, d b^-}{(2
    \pi)^3} \Tr \left[ W (p,b) \right] =1,
\end{align}
such that 
\begin{align}
\label{Wunp_norm}
g_{+-} \int \frac{d^{2} p_\perp \, d p^+ \, d^2 b_\perp \, d b^-}{(2
  \pi)^3} \, W_{unp} (p,b) = \half .
\end{align}

\vspace{0.5cm}

In a similar way, the quark correlator \eqref{e:qcorr1} of the nucleon
can be expanded in a basis of spin states, and the polarized and
unpolarized parts can be projected using the Pauli matrices:
\begin{align} 
  \label{e:qcorr2}
  \phi(x, \ul{k}; S) &= \phi_{unp}(x, \ul{k}) - S_\mu
  \hat{\phi}_{pol}^\mu (x, \ul{k})
  \notag \\
  \phi_{\lambda \lambda'} (x, \ul{k}) &= \phi_{unp} (x, \ul{k}) \:
  \bigg[ \:\: \mathbf{1} \:\: \bigg]_{\lambda \lambda'} -
  \hat{\phi}_{pol, \, \mu} (x, \ul{k}) \: \bigg[ \: \hat\sigma^\mu \:
  \bigg]_{\lambda \lambda'},
\end{align}
where
\begin{align}
  \phi_{unp} (x, \ul{k}) &= \half \sum_{\lambda \lambda'} \bigg[
  \phi_{\lambda \lambda'}(x, \ul{k}) \: \mathbf{1}_{\lambda' \lambda}
  \bigg] = \half \Tr[ \phi(x, \ul{k}) ]
 \notag \\
 \hat{\phi}_{pol}^\mu (x, \ul{k}) &= \half \sum_{\lambda \lambda'}
 \bigg[ \phi_{\lambda \lambda'}(x, \ul{k}) \: \hat\sigma^\mu_{\lambda'
   \lambda} \bigg] = \half \Tr[ \phi(x, \ul{k}) \: \hat\sigma^\mu],
\end{align}
and the matrix form is
\begin{align}
  \phi_{\lambda \lambda'} (x, \ul{k}) &\equiv \frac{\gpm}{(2\pi)^3} \,
  \int d^{2-}r \, e^{i k \cdot r} \, \bra{N (P,\lambda)} \,
  \psibar_\beta (0) \, \mathcal{U}[0,r] \, \psi_\alpha (r) \ket{N (P,
    \lambda')}.
\end{align}
Note that $S_\mu$ and $\hat\sigma^\mu$ above depend on the (averaged)
momentum of the nucleon $\bar p$, which is not shown explicitly. The
expansion \eqref{e:qTMD1} of the quark correlator into the 8
leading-twist TMDs is defined in terms of the longitudinal spin $S_L$
and the transverse spin $\ul{S}$.  The Dirac projections of the
correlator $\phi^{[\Gamma]}$ as in \eqref{e:qTMD2} are
boost-invariant, so we can evaluate them in the rest frame in which
$\vec{S} = (S_\bot^1 , S_\bot^2 , S_L)$; this allows us to identify
the projections $\phi_{unp}^{[\Gamma]}$ and $\hat\phi_{pol}^{\mu \,
  [\Gamma]}$ in terms of the TMDs as
\begin{align} \label{e:qTMD3} \phi_{unp}^{[\Gamma]} & \,= \bigg( f_1
  \bigg) \bigg[ \frac{1}{4} \gpm \Tr [ \gamma^- \Gamma ] \bigg] +
  \bigg( \frac{k_\bot^i}{m} h_1^\bot \bigg) \bigg[ \frac{i}{4} \gpm
  \Tr [ \gamma_{\bot i} \gamma^- \Gamma ] \bigg]
  \notag \\
  \phi_{pol}^{0 \, [\Gamma]} &\rfeq 0 \notag \\
  \phi_{pol}^{3 \, [\Gamma]} &\rfeq \bigg( g_1 \bigg) \bigg[
  \frac{1}{4} \gpm \Tr [\gamma^5 \gamma^- \Gamma] \bigg] + \bigg(
  \frac{k_\bot^i}{m} h_{1 L}^\bot \bigg) \bigg[ \frac{1}{4} \gpm \Tr[
  \gamma^5 \gamma_{\bot i} \gamma^- \Gamma] \bigg]
  \notag \\
  \phi_{pol, \, \bot}^{j \, [\Gamma]} &\rfeq \bigg( -
  \frac{k_\bot^i}{m} \epsilon_T^{i j} f_{1T}^\bot \bigg) \bigg[
  \frac{1}{4} \gpm \Tr[ \gamma^- \Gamma] \bigg] + \bigg(
  \frac{k_\bot^j}{m} g_{1T} \bigg) \bigg[ \frac{1}{4} \gpm \Tr[
  \gamma^5 \gamma^- \Gamma] \bigg]
  \notag \\ & \hspace{1cm} +
  \bigg( \delta^{i j} h_{1T} + \frac{k_\bot^i k_\bot^j}{m^2}
  h_{1T}^\bot \bigg) \bigg[ \frac{1}{4} \gpm \Tr[ \gamma^5
  \gamma_{\bot i} \gamma^- \Gamma] \bigg].
\end{align}

\begin{figure}[hbt]
 \centering
 \includegraphics[width=\textwidth]{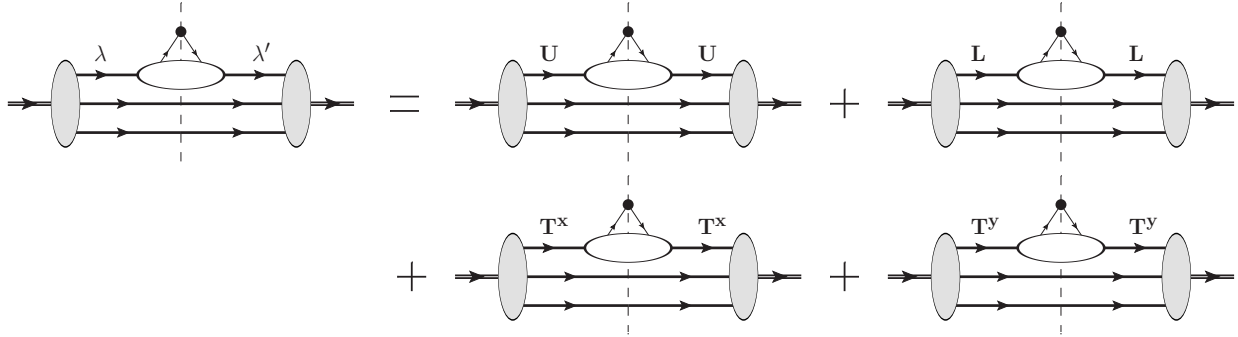}
 \caption{Schematic decomposition of the spin sum occurring in
   \eqref{e:QCFact3}.  Summing over the 4 independent components of
   the $( 2 \times 2 )$ matrices $[\phi]_{\lambda \lambda'} , \,
   [W]_{\lambda' \lambda}$ corresponds to summing over the 4 possible
   intermediate polarizations of the nucleons: $U$ = unpolarized, $L$
   = longitudinally polarized, $T^j$ = transversely polarized in the
   $\hat{j}$ direction.}
 \label{f:Spin_Sum} 
\end{figure}

In the derivation of \eqref{e:QCFact2}, a sum over the spins of the
intermediate nucleons occurs independently in the amplitude and
complex-conjugate amplitude (see Fig.~\ref{f:QCFact}).  Thus the spins
entering the quasi-classical factorization formula are necessarily
non-diagonal, which can be compactly expressed in terms of the
matrices $W_{\lambda \lambda'}$ and $\phi_{\lambda \lambda'}$:
\begin{align}
  \Phi^A (x, \ul{k}; P) &= A \frac{g_{+-}}{(2\pi)^5} \, \sum_{\lambda
    \lambda'} \, \int d^{2+} p \, d^{2-} b \, d^2 r \, d^2 k' \, e^{-i
    (\ul{k} - \ul{k'} - \hat{x} \ul{p}) \cdot \ul{r}}
  \notag \\ & \hspace{2.5cm} \times \:\:
  W_{\lambda \lambda'} (p,b; P) \, \phi_{\lambda' \lambda}^N (\hat{x},
  \ul{k}' ; p) \, S_{(r_T, b_T)}^{[\infty^-, b^-]}
  \notag \\ &=
  A \frac{g_{+-}}{(2\pi)^5} \, \int d^{2+} p \, d^{2-} b \, d^2 r \,
  d^2 k' \, e^{-i (\ul{k} - \ul{k'} - \hat{x} \ul{p}) \cdot \ul{r}} \,
  \notag \\ & \hspace{2.5cm} \times \:\:
  \Tr \bigg[ W (p,b; P) \: \phi(\hat{x}, \ul{k}' ; p) \bigg] \,
  S_{(r_T, b_T)}^{[\infty^-, b^-]},
\end{align}
where we have restricted the discussion to an unpolarized nucleus.
From \eqref{e:Wig5}, \eqref{e:spinRF4}, and \eqref{e:qcorr2}, the
trace of the two matrices gives
\begin{align}
  \half \Tr \bigg[ W \, \cdot \, \phi \bigg] &= W_{unp} \, \phi_{unp}
  - \hat{W}_{pol , \, \mu} \, \hat\phi_{pol}^\mu
  \notag \\ & \rfeq
  W_{unp} \, \phi_{unp} +  \vec{W}_{pol} \cdot \vec\phi_{pol},
\end{align}
which sums over all four independent spin states of the intermediate
nucleons.  Thus we recover a form similar to \eqref{e:QCFact2} in
which the spins of the nucleons are diagonal (i.e., the same in the
amplitude and complex-conjugate amplitude),
\begin{align} 
  \label{e:QCFact3}
  \Phi^A (x, \ul{k}; P) &= \frac{2 A \, g_{+-}}{(2\pi)^5} \, \int
  d^{2+} p \, d^{2-} b \, d^2 r \, d^2 k' \, e^{-i (\ul{k} - \ul{k'} -
    \hat{x} \ul{p}) \cdot \ul{r}}
  \notag \\ & \times \:\:
  \bigg( W_{unp} (p,b; P) \, \phi_{unp} (\hat{x}, \ul{k}' ; p) -
  \hat{W}_{pol , \, \mu} (p,b; P) \hat\phi_{pol}^\mu (\hat{x}, \ul{k}'
  ; p) \bigg) \, S_{(r_T, b_T)}^{[\infty^-, b^-]},
\end{align}
as illustrated in Fig.~\ref{f:Spin_Sum}.  The sum now effectively runs
over the four polarizations: $U$ = unpolarized, $L$ =
longitudinally-polarized, and $T^j$ = transversely polarized in the
$\hat{j}$ direction.  This formulation makes the projections defined
by \eqref{e:Wig3}, \eqref{e:qcorr2} independent of the choice of spin
basis and expresses the spin dependence in a manifestly
Lorentz-covariant form.

%***********************************************************************************
%***********************************************************************************

%|||||||||||||||||||||||||||||||||||||||||||||||||||||||||||||||||||||||||||||||||||

\subsection{Parameterization of the Wigner Distribution}
\label{sec:paramet}
%|||||||||||||||||||||||||||||||||||||||||||||||||||||||||||||||||||||||||||||||||||

The quasi-classical factorization formula \eqref{e:QCFact3} describes
how the properties of the Wigner distribution and the multiple
rescattering build up the TMDs of the nucleus from the various TMDs
of the nucleons.  One powerful feature of \eqref{e:QCFact3} is that
the Wigner distribution is constructed purely from the light-front
wave functions (see, e.g. \eqref{e:Wigmat}), without contamination
from the scattering dynamics embodied in the gauge link.  It is
therefore invariant under several symmetries which the TMDs
themselves are not, such as $PT$.  We would like to use all the
applicable symmetries to parameterize the types of structures which
can occur in the Wigner distribution.

The Wigner distribution also does not know about the direction of the
collision axis; that information enters operatorially through the
gauge link $\mathcal{U}[0,r]$ or diagrammatically through the
interaction with the virtual photon.  We may choose to describe the
functional dependence of the Wigner distribution using variables which
are appropriate for a collision along the $z$-axis (e.g. $p^+,
\ul{p}$), but the distribution itself should not treat $z$ as a
``special'' direction.  Naively, one would expect that, when viewed
from the rest frame of the nucleus, the Wigner distribution $W(p,b)$
should possess {\it three}-dimensional rotation symmetry due to the
lack of a preferred collision axis.  This is in contrast to the
factors $\phi (x, \ul{k}) $ and $S (r_T, b_T)$ containing the gauge
link, which have at most {\it two}-dimensional rotational invariance
due to the special role of the longitudinal direction.

There are some subtleties about the application of full rotational
symmetry to the Wigner distribution constructed from light-front wave
functions; these are addressed most clearly using the ``covariant
light-front formalism'' of \cite{Carbonell:1998rj}.  We discuss these
issues in Appendix~\ref{sec:Karmanov}.  The conclusion is that, if the
nucleons move nonrelativistically in the nucleus, then their wave
functions -- and hence the Wigner distribution in \eqref{e:Wig7} and
\eqref{e:Wig8} -- do possess the desired rotational invariance in the
nuclear rest frame.  The boost-friendly variables $\alpha , b^-$ are
given in terms of the rotation-friendly variables $\vec{p}, \vec{b}$
through \eqref{e:NRvars}.  We can now take advantage of this symmetry
to constrain the functional form of the Wigner distribution.

%-----------------------------------------------------------------------------------

\subsubsection{Wigner Distribution of an Unpolarized Nucleus}

%-----------------------------------------------------------------------------------

For an unpolarized nucleus with a non-relativistic distribution of
nucleons, the Wigner distribution \eqref{e:Wig8} possesses manifest
rotational invariance in the rest frame and depends on the nucleon
spin only linearly.  Moreover, because the Wigner distribution is
built purely from the nucleon wave functions, it possesses the
unbroken discrete symmetries of (ordinary) parity and time reversal.
The $P$ and $T$ symmetries of the wave functions translate into the
Wigner distribution in the expected way:
\begin{align}
  W(\vec{p} , \vec{b} , \vec{S}) \overset{P}{=} W(- \vec{p} , -
  \vec{b}, \vec{S}) \overset{T}{=} W(-\vec{p} , \vec{b} , -\vec{S}).
\end{align}
We can therefore constrain the form of the Wigner distribution using
these discrete symmetries and rotational invariance.

For the unpolarized distribution, requiring rotational invariance
yields
\begin{align}
  W_{unp} (\vec{p} , \vec{b}) = W_{unp} \left[ \vec{p}^2 , \vec{b}^2 ,
    \vec{p} \cdot \vec{b} \right],
\end{align}
but the quantity $\vec{p} \cdot \vec{b}$ is odd under time reversal,
so $W_{unp}$ can only depend on it through even powers:
\begin{align}
  W_{unp} (\vec{p} , \vec{b}) = W_{unp} \left[ \vec{p}^2 , \vec{b}^2 ,
    (\vec{p} \cdot \vec{b})^2 \right].
\end{align}
Now let us change variables to the boost-invariant quantities using
\eqref{e:NRvars}:
\begin{align} \label{e:NRvars2}
  \vec{p}^2 &= p_T^2 + (\alpha - \tfrac{1}{A})^2 M_A^2 \notag \\
  \vec{b}^2 &= b_T^2 + (\tfrac{1}{M_A})^2 (\gpm P^+ b^-)^2 \notag \\
  (\vec{p} \cdot \vec{b})^2 &= \left[ \ul{p} \cdot \ul{b} - (\alpha -
    \tfrac{1}{A}) (\gpm P^+ b^-) \right]^2
  \notag \\ &=
  (\ul{p} \cdot \ul{b})_T^2 + (\alpha - \tfrac{1}{A})^2 (\gpm P^+ b^-)^2 -
  2 (\ul{p} \cdot \ul{b}) (\alpha - \tfrac{1}{A}) (\gpm P^+ b^-),
\end{align}
where $\alpha = p^+ / P^+$ is the longitudinal momentum fraction of
the nucleon in the nucleus.  In terms of these quantities, the
unpolarized distribution depends on
\begin{align}
  W_{unp} (\alpha, \ul{p} \, ; b^- , \ul{b}) = W_{unp} \left[ p_T^2 , b_T^2
    , (\ul{p} \cdot \ul{b})_T^2 \, ; \, (\alpha - \tfrac{1}{A})^2 , (P^+
    b^-)^2 \, ; \, (\ul{p} \cdot \ul{b}) (\alpha-\tfrac{1}{A})(P^+ b^-)
  \right],
\end{align}
where we have separated out the dependence on the longitudinal and
transverse variables.

There is one interesting term which appears to mix the longitudinal
and transverse variables, arising from the cross term $(p_\parallel
b_\parallel) (\ul{p} \cdot \ul{b})$ of $(\vec{p} \cdot \vec{b})^2$.
In principle such a term is not prohibited by the symmetries of Wigner
distribution; however, looking back at the quasi-classical
factorization formula \eqref{e:QCFact3}, we see that this term will
not contribute to the cross-section.  The reason is that the only
other dependence on $\vec{b}$ arises from the dipole scattering
amplitude $S_{(r_T , b_T)}^{[\infty^- , b^-]}$ \eqref{e:GGM}.  This
factor does not possess the same three-dimensional rotation invariance
as the Wigner distribution \eqref{e:Wig8}, because the multiple
scattering depends on the depth $b^-$ of the struck nucleon.  It does,
however, retain a residual two-dimensional rotation invariance in the
transverse plane, depending only on $b^- , b_T$.  Thus a term like $
\int d^2 b \, (p_\parallel b_\parallel) (\ul{p} \cdot \ul{b}) \,
f(b_T) $ would integrate out to zero in \eqref{e:QCFact3} because of
antisymmetry under $\ul{b} \rightarrow - \ul{b}$.  Therefore the terms
of the Wigner distribution which survive the $d^2 b$ integral can only
depend on the square of this cross-term, $(\ul{p} \cdot \ul{b})_T^2
(\alpha-1/A)^2 (P^+ b^-)^2 $, and the dependence on these squared
factors are already taken into account.  Thus we can replace
\begin{align}
  W_{unp} (\alpha, \ul{p} \, ; b^- , \ul{b}) \Rightarrow W_{unp} \left[
    p_T^2 , b_T^2 , (\ul{p} \cdot \ul{b})_T^2 ; (\alpha - \tfrac{1}{A})^2 ,
    (P^+ b^-)^2 \right]
\end{align}
without missing out on any terms which would survive the $d^2 b$
integral in \eqref{e:QCFact3}.  We use an arrow here rather than an
equality to emphasize that we are now dropping contributions to the
Wigner distribution which are permitted by symmetry, but would not
survive to contribute to the nuclear TMD.

In fact, we can carry this argument one step farther: after the
integration
\begin{align} 
  \notag \int d^2 b \, W_{unp} (\alpha, \ul{p} \, ; b^- , \ul{b}) \, S(r_T
  , b_T ; b^-)
\end{align}
is performed, all sensitivity to the direction of the two-vector
$\ul{b}$ drops out.  Thus we can replace
\begin{align}
  b_\bot^i b_\bot^j \rightarrow \half b_T^2 \delta^{i j}
\end{align}
in the Wigner distribution now, without loss of generality in the
types of terms which can contribute to the quasi-classical
factorization formula \eqref{e:QCFact3}.  Doing so further reduces the
number of terms which can contribute, since
\begin{align}
  (\ul{p} \cdot \ul{b})_T^2 = p_\bot^i p_\bot^j \, (b_\bot^i b_\bot^j)
  \rightarrow \half p_T^2 b_T^2.
\end{align} 
Thus, from all of these simplifications, we can write the relevant
part of the unpolarized Wigner distribution as
\begin{align} \label{e:Wunp} W_{unp} (\alpha, \ul{p} \, ; b^- , \ul{b})
  \Rightarrow W_{unp} \left[ p_T^2 , b_T^2 ; (\alpha - \tfrac{1}{A})^2 ,
    (P^+ b^-)^2 \right].
\end{align}
By considering the constraints due to parity, time reversal, and
three-dimensional rotation invariance of the Wigner distribution,
together with the two-dimensional rotational invariance of the
multiple scattering factor (gauge link) to which it couples, we have
reduced the unpolarized distribution down to a form which is
manifestly even under $\ul{p} \rightarrow - \ul{p},$ $\ul{b}
\rightarrow - \ul{b},$ $(\alpha- \tfrac{1}{A}) \rightarrow - (\alpha -
\tfrac{1}{A}),$ and $b^- \rightarrow - b^-$.  These symmetries
strongly constrain the interplay between the factors $W(p,b) , \phi
(x, k') , S(r, b)$ entering the quasi-classical factorization formula
\eqref{e:QCFact3}.

Similarly, we can impose rotational invariance on the polarized part
of the Wigner distribution:
\begin{align}
  \vec{S} \cdot \vec{W}_{pol} (\vec{p} , \vec{b}) &= \left( \vec{S}
    \cdot \vec{b} \, \right) \: W_1 \! \left[\vec{p}^2 , \vec{b}^2 ,
    \vec{p} \cdot \vec{b} \right] + \left( \vec{S} \cdot \vec{p} \,
  \right) \: W_2 \! \left[ \vec{p}^2 , \vec{b}^2 , \vec{p} \cdot
    \vec{b} \right]
  \notag \\ & \hspace{2cm} + 
  \left( \vec{S} \cdot( \vec{b} \times \vec{p} ) \right) W_3 \! \left[
    \vec{p}^2 , \vec{b}^2 , \vec{p} \cdot \vec{b} \right].
\end{align}
The spin-dependent factors $(\vec{S} \cdot \vec{b})$ and $(\vec{S}
\cdot \vec{p})$ are odd under parity, while all of the arguments
$\vec{p}^2 , \vec{b}^2 , \vec{p} \cdot \vec{b}$ are $P$-even;
therefore these types of spin dependence cannot enter into the
polarized Wigner distribution.  The only spin dependence which is
permitted by parity is the factor $(\vec{S} \cdot \vec{L})$ describing
spin-orbit coupling, where $\vec{L} = \vec{b} \times \vec{p}$ is the
orbital angular momentum of the nucleon.  Since $(\vec{S} \cdot
\vec{L})$ is even under both parity and time reversal, the $T$-odd
quantity $(\vec{p} \cdot \vec{b})$ can only occur in even powers, just
like in the unpolarized distribution:
\begin{align}
  \vec{S} \cdot \vec{W}_{pol} (\vec{p} , \vec{b}) &= \left( \vec{S}
    \cdot( \vec{b} \times \vec{p} ) \right) W_3 \! \left[ \vec{p}^2 ,
    \vec{b}^2 , (\vec{p} \cdot \vec{b})^2 \right].
\end{align}
Now let us again change back to the boost-invariant variables
\eqref{e:NRvars}.  In addition to the expressions in
\eqref{e:NRvars2}, we must account for the spin-dependent factor
\begin{align}
  \vec{S} \cdot (\vec{b} \times \vec{p}) &= S_\parallel (\ul{b} \times
  \ul{p}) + b_\parallel (\ul{p} \times \ul{S}) + p_\parallel (\ul{S}
  \times \ul{b})
  \notag \\ &=
  \lambda \, (\ul{b} \times \ul{p}) - \tfrac{1}{M_A} (\gpm P^+ b^-)
  \,(\ul{p} \times \ul{S}) + (\alpha - \tfrac{1}{A}) M_A \, (\ul{S} \times
  \ul{b})
\end{align}
to write
\begin{align}
  - S_\mu \hat{W}_{pol}^\mu (\alpha, \ul{p} \, ; b^- , \ul{b}) &= \left(
    \lambda \, (\ul{b} \times \ul{p}) - \tfrac{1}{M_A} (\gpm P^+ b^-)
    \,(\ul{p} \times \ul{S}) + (\alpha - \tfrac{1}{A}) M_A \, (\ul{S}
    \times \ul{b}) \right)
  \notag \\ &\times
  W_{3} \left[ p_T^2 , b_T^2 , (\ul{p} \cdot \ul{b})_T^2 \, ; \, (\alpha -
    \tfrac{1}{A})^2 , (P^+ b^-)^2 \, ; \, (\ul{p} \cdot \ul{b})
    (\alpha-\tfrac{1}{A})(P^+ b^-) \right].
\end{align}
As before, we now replace $b_\bot^i b_\bot^j \rightarrow \thalf b_T^2
\delta^{i j}$ to keep only the terms which can contribute to the
nuclear TMD.  In addition to reducing $(\ul{p} \cdot \ul{b})_T^2
\rightarrow \thalf p_T^2 b_T^2$, this also simplifies the spin
dependence for the factors which couple $\ul{S}$ to $\ul{b}$:
\begin{align}
  \left[ \lambda \, (\ul{b} \times \ul{p}) \right] \left[ (\ul{p}
    \cdot \ul{b}) (\alpha-\tfrac{1}{A})(P^+ b^-) \right] &\rightarrow
  \thalf \lambda (\alpha - \tfrac{1}{A}) (P^+ b^-) \:\: (\ul{p} \times
  \ul{p}) = 0
  \notag \\ 
  \left[ (\alpha - \tfrac{1}{A}) \, (\ul{S} \times \ul{b}) \right] \left[
    (\ul{p} \cdot \ul{b}) (\alpha-\tfrac{1}{A})(P^+ b^-) \right]
  &\rightarrow - \thalf (\alpha - \tfrac{1}{A})^2 \:\: (P^+ b^-) (\ul{p}
  \times \ul{S}).
\end{align}
We see that the spin dependence of the form $\lambda (\ul{b} \times
\ul{p})$ has dropped out completely, and the spin dependence of the
form $(\alpha - \tfrac{1}{A}) (\ul{S} \times \ul{b})$ has reduced down
to the form $(P^+ b^-) (\ul{p} \times \ul{S})$.  Thus, without loss of
generality, we can identify the structure $(P^+ b^-) (\ul{p} \times
\ul{S})$ as the only spin dependence which survives the $d^2 b$
integral to contribute to the nuclear TMDs:
\begin{align} \label{e:Wpol} - S_\mu \hat{W}_{pol}^\mu (\alpha, \ul{p} \, ;
  b^- , \ul{b}) &\Rightarrow \left(\frac{\gpm P^+ b^-}{M_A}\right)
  (\ul{p} \times \ul{S}) \: W_{OAM} \! \left[ p_T^2 , b_T^2 \, ; \,
    (\alpha-\tfrac{1}{A})^2 , (P^+ b^-)^2 \right] .
\end{align}
Again, consideration of the relevant symmetries has reduced the form
of the polarized Wigner distribution significantly.  The structure
that survives contains a prefactor linear in the transverse spin which
is odd under $b^- \rightarrow - b^-$ and $\ul{p} \rightarrow -
\ul{p}$, times a function which is manifestly even under $\ul{p}
\rightarrow - \ul{p},$ $\ul{b} \rightarrow - \ul{b},$
$(\alpha-\tfrac{1}{A}) \rightarrow - (\alpha - \tfrac{1}{A}),$ and
$b^- \rightarrow - b^-$.  The designation ``OAM'' reflects the fact
that this structure originated from the presence of $\vec{L} \cdot
\vec{S}$ coupling in the rest frame.

%
%***********************************************************************************
\begin{figure}[bt]
 \centering
 \includegraphics[width= \textwidth]{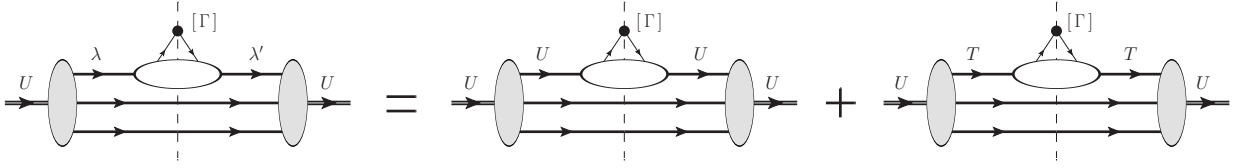}
 \caption{Spin structure of the unpolarized nucleus.  An unpolarized
   nucleus can give rise to unpolarized nucleons or
   transversely-polarized nucleons, which contribute to the quark
   distribution through various nucleonic TMDs.  Interestingly there
   is no contribution from longitudinally-polarized nucleons in the
   unpolarized nucleus \eqref{e:Wig9}.}
\label{f:Spin_Sum2}
\end{figure}
%***********************************************************************************
%

Altogether, \eqref{e:Wunp} and \eqref{e:Wpol} allow us to parameterize
the most general structure of the Wigner distribution which is
consistent with parity, time reversal, and 2D and 3D rotation
symmetries.  The structure which can contribute to the quark TMD of an
unpolarized nucleus is
\begin{align}
  W(\alpha, \ul{p} \, ; b^- , \ul{b} \, ; \lambda , \ul{S}) \: \Rightarrow
  \:\: &W_{unp} \left[ p_T^2 , b_T^2 ; (\alpha - \tfrac{1}{A})^2 , (P^+
    b^-)^2 \right]
  \notag \\ &+
  \left(\tfrac{\gpm P^+ b^-}{M_A}\right) (\ul{p} \times \ul{S}) \:\:
  W_{OAM} \! \left[ p_T^2 , b_T^2 \, ; \, (\alpha-\tfrac{1}{A})^2 , (P^+ b^-)^2 \right],
  \label{W_param}
\end{align}
or, in the notation of \eqref{e:Wig3},
\begin{align} 
  \label{e:Wig9}
  W_{unp} (p, b)  & \, = W_{unp} \left[ p_T^2 , b_T^2 ; (\alpha - \tfrac{1}{A})^2 , (P^+ b^-)^2 \right] \notag \\
  \hat{W}_{pol}^0 (p, b) &\rfeq 0 \notag \\
  \hat{W}_{pol}^3 (p, b) &\rfeq 0 \notag \\
  \hat{W}_{pol, \, \bot}^j (p, b) &\rfeq \left(\tfrac{\gpm P^+
      b^-}{M_A}\right) p_\bot^i \epsilon_T^{i j} W_{OAM} \! \left[
    p_T^2 , b_T^2 \, ; \, (\alpha-\tfrac{1}{A})^2 , (P^+ b^-)^2
  \right].
\end{align}
\eq{W_param} is illustrated in \fig{f:Spin_Sum2}. The parameterization
given here was derived for the distribution of nucleons with
polarization $\vec{S}_N = (\ul{S}, \lambda)$ moving
nonrelativistically in an unpolarized nucleus, for which the only
nontrivial spin dependence entered as $(\vec{L} \cdot \vec{S}_N)$.  It
is straightforward to extend this parameterization to describe a
polarized nucleus as well.  In addition to the equivalent spin-orbit
coupling $(\vec{L} \cdot \vec{S}_A)$ which was the source of the
``orbital angular momentum channel'' to generate the nuclear Sivers
function $f_{1T}^{\bot A}$ in \cite{Kovchegov:2013cva}, there are
spin-spin couplings like $(\vec{S}_N \cdot \vec{S}_A)$ and
spin-orbit-spin couplings like $(\vec{S}_N \cdot \vec{p}) (\vec{p}
\cdot \vec{S}_A)$.  We will leave this generalization to the full
structure of the Wigner distribution of a polarized nucleus for future
work.

%|||||||||||||||||||||||||||||||||||||||||||||||||||||||||||||||||||||||||||||||||||

\subsection{Quasi-Classical TMDs of an Unpolarized Nucleus}
\label{sec-unpnuc}

%|||||||||||||||||||||||||||||||||||||||||||||||||||||||||||||||||||||||||||||||||||

The TMD quark correlator $\Phi_{unp}^A$ of an unpolarized nucleus from
\eqref{e:qTMD3} contains just two leading-twist TMDs: the unpolarized quark
distribution $f_1^A$ and the Boer-Mulders distribution $h_1^{\bot A}$:
\begin{align} 
  \label{e:unp2}
  \Phi_{unp}^{A \, [\Gamma]} (x,\ul{k}) = \bigg( f_1^A (x, k_T) \bigg)
  \bigg[ \frac{1}{4} \gpm \Tr [ \gamma^- \Gamma ] \bigg] + \bigg(
  \frac{k_\bot^i}{M_A} h_1^{\bot A} (x, k_T) \bigg) \bigg[ \frac{i}{4}
  \gpm \Tr [ \gamma_{\bot i} \gamma^- \Gamma ] \bigg]
\end{align}
Because $f_1^A$ corresponds to unpolarized quarks and $h_1^{\bot A}$
corresponds to transversely-polarized quarks, these two TMDs can be
projected from the correlator as in \eqref{e:qTMD2} by tracing the
Dirac structure with $\Gamma = \gamma^+$ or $\Gamma = \gamma^5
\gamma^+ \gamma_\bot^j$, respectively:
\begin{align} 
  \label{e:unp3}
  f_1^A (x, k_T) &=\Phi_{unp}^{A \, [\gamma^+]} (x, \ul{k})
  % = \half \left[\Phi_{unp}^{A \, [\gamma^+]} (x, \ul{k}) + (\ul{k}
  %   \rightarrow - \ul{k})\right]
 %
 \notag \\
 \epsilon_T^{j i} \frac{k_\bot^i}{M_A} h_1^{\bot A} (x, k_T) &=
 \Phi_{unp}^{A \, [\gamma^5 \gamma^+ \gamma_\bot^j]} (x, \ul{k})
 % = \half \left[ \Phi_{unp}^{A \, [\gamma^5 \gamma^+ \gamma_\bot^j]}
 %   (x, \ul{k}) - (\ul{k} \rightarrow - \ul{k})\right] .
\end{align}
These nuclear TMDs are related to the TMDs of the nucleons through
the quasi-classical factorization formula
\begin{align} 
  \label{e:QCFact4}
  \Phi_{unp}^{A \, [\Gamma]} (x, \ul{k}) &= \frac{2 A \,
    g_{+-}}{(2\pi)^5} \, \int d^{2+} p \, d^{2-} b \, d^2 r \, d^2 k'
  \, e^{-i (\ul{k} - \ul{k'} - \hat{x} \ul{p}) \cdot \ul{r}}
  \notag \\ & \times \:\:
  \bigg( W_{unp} (p,b) \, \phi_{unp}^{[\Gamma]} (\hat{x}, \ul{k}' ) -
  \hat{W}_{pol , \, \mu} (p,b) \hat\phi_{pol}^{\mu \, [\Gamma]}
  (\hat{x}, \ul{k}' ) \bigg) \, S_{(r_T, b_T)}^{[\infty^-, b^-]},
\end{align}
with the Wigner distribution of nucleons parameterized by
\eqref{e:Wig9}.  Since the nucleons can be polarized, the active
quarks can come from one of several nucleonic TMDs depending on the
polarization, as in \eqref{e:qTMD3}.  Combining the Wigner
distribution for a given nucleon polarization with the associated
TMDs, we obtain
\begin{align} 
  \label{e:spinorb1}
  W_{unp} & (p, b) \, \phi_{unp}^{[\Gamma]} (\hat{x}, \ul{k}') =
  W_{unp} \left[ p_T^2 , b_T^2 ; (\alpha-\tfrac{1}{A})^2, (P^+ b^-)^2
  \right]
  \notag \\  &\times
  \bigg\{ \bigg( f_1^N (\hat{x}, k_T^\prime) \bigg) \bigg[ \frac{1}{4}
  \gpm \Tr [ \gamma^- \Gamma ] \bigg] + \bigg( \frac{k_\bot^{\prime \,
      i}}{m_N} h_1^{\bot N} (\hat{x}, k_T^\prime) \bigg) \bigg[
  \frac{i}{4} \gpm \Tr [ \gamma_{\bot i} \gamma^- \Gamma ] \bigg]
  \bigg\}
  \notag \\ \, \notag \\
  - \hat{W}_{pol, \mu} & (p,b) \, \hat{\phi}_{pol}^{\mu \, [\Gamma]}
  (\hat{x}, \ul{k}')= +\hat{W}_{pol, \bot}^j (p,b) \, \hat{\phi}_{pol,
    \, \bot}^{j \, [\Gamma]} (\hat{x}, \ul{k}')
  \notag \\ &=
  \left(\tfrac{\gpm P^+ b^-}{M_A}\right) p_\bot^\ell \epsilon_T^{\ell
    j} \, W_{OAM} \! \left[ p_T^2 , b_T^2 \, ; \, (\alpha-\tfrac{1}{A})^2 ,
    (P^+ b^-)^2 \right]
  \notag \\ &\times \bigg\{
  \bigg( - \frac{k_\bot^{\prime \, i}}{m_N} \epsilon_T^{i j}
  f_{1T}^{\bot N} (\hat{x}, k_T^\prime) \bigg) \bigg[ \frac{1}{4} \gpm
  \Tr[ \gamma^- \Gamma] \bigg] + \bigg( \frac{k_\bot^{\prime \, j}}{m_N}
  g_{1T}^{\bot N} (\hat{x}, k_T^\prime) \bigg) \bigg[ \frac{1}{4} \gpm
  \Tr[ \gamma^5 \gamma^- \Gamma] \bigg]
  \notag \\ & \hspace{1cm} +
  \bigg( \delta^{i j} h_{1T}^N (\hat{x}, k_T^\prime) +
  \frac{k_\bot^{\prime \, i} k_\bot^{\prime \, j}}{m_N^2} h_{1T}^{\bot
    N} (\hat{x}, k_T^\prime) \bigg) \bigg[ \frac{1}{4} \gpm \Tr[
  \gamma^5 \gamma_{\bot i} \gamma^- \Gamma] \bigg] \bigg\},
\end{align}
where $\alpha = p^+ / P^+$ and $\hat{x} = x/\alpha$.
The projection $[\Gamma]$ of the nuclear TMD quark distribution
$\Phi^A$ in \eqref{e:unp3} selects the distribution of quarks with a
given polarization $(U, L, T^j)$.  This projection also applies to the
nucleonic TMD quark distributions $\phi$ in \eqref{e:spinorb1},
picking out a specific combination of nucleonic TMDs which can
contribute to a given TMD of the nucleus.  The orbital dependence on
the quark momentum $\ul{k}'$ carried by each of the TMDs can couple
to the orbital dependence on the nucleon momentum $\ul{p}$ in the
Wigner distribution, giving rise to nontrivial interplay between the
TMDs of the nucleons, the orbital angular momentum these nucleons
carry within the nucleus, and the resulting TMDs of the nucleus.  Now
let us use these properties to separately study the quasi-classical
decomposition of the unpolarized quark distribution $f_1^A$ and the
Boer-Mulders distribution $h_1^{\bot A}$ of the nucleus.

%-----------------------------------------------------------------------------------

\subsubsection{The Unpolarized Quark Distribution $f_1^A$}

%-----------------------------------------------------------------------------------

The projection $\Gamma = \gamma^+$ in \eqref{e:unp3} selects out the
unpolarized quark distribution $f_1^A$ in the nucleus and the
unpolarized quark distribution $f_1^N$ and quark Sivers function
$f_{1T}^{\bot N}$ of the nucleon in \eqref{e:spinorb1} (see also
\eqref{e:qTMD2}) .  Using this in \eqref{e:QCFact4} gives
\begin{align} 
  \label{e:f1-1}
  f_1^A (x, k_T) &= \frac{2 A \, g_{+-}}{(2\pi)^5} \, \int d^{2+} p \,
  d^{2-} b \, d^2 r \, d^2 k' \, e^{-i (\ul{k} - \ul{k'} - \hat{x}
    \ul{p}) \cdot \ul{r}} \, S_{(r_T, b_T)}^{[\infty^-, b^-]}
   \notag \\ & \times \:\:
   \bigg( W_{unp} (p,b) \, f_1^N (\hat{x}, k_T^\prime) -
   \tfrac{\gpm}{M_A m_N} (P^+ b^-) (\ul{p} \cdot \ul{k}') \, W_{OAM}
   (p,b) \, f_{1T}^{\bot N} (\hat{x}, k_T^\prime) \bigg) \,,
\end{align}
which shows that $(\vec{L} \cdot \vec{S}_N)$ spin-orbit coupling given
by $W_{OAM}$ can result in a mixing between the unpolarized quark
distribution $f_1^A$ of the nucleus and the quark Sivers function
$f_{1T}^{\bot N}$ of the nucleons.  This is the mirror effect to the
OAM channel that was found in \cite{Kovchegov:2013cva}, in which
$(\vec{L} \cdot \vec{S}_A)$ coupling in a polarized nucleus gave rise
to a mixing between the Sivers function $f_{1T}^{\bot A}$ of the
nucleus and the unpolarized $f_1^N$ distribution of the nucleons.

The unpolarized quark distribution $f_1$ is a $PT$-even function and
therefore process independent (see \cite{Collins:1992kk} and others);
that is, it should give the same result when calculated with a
future-pointing gauge link like \eqref{e:GGM} appropriate for SIDIS as
when calculated with a past-pointing gauge link like
\begin{align} 
  \label{e:GGM2}
  S_{(r_T , b_T)}^{[b^- , -\infty^-]} = \exp\left[ -\frac{1}{4} r_T^2
    \, Q_s^2 (b_T) \left(\frac{b^- + R^- (b_T)}{2 R^- (b_T)} \right)
    \ln\frac{1}{r_T \Lambda} \right]
\end{align}
appropriate for the Drell-Yan process (DY).  In the quasi-classical
approximation, the difference between the future-pointing gauge link
of \eqref{e:GGM} and the past-pointing gauge link of \eqref{e:GGM2} is
in the fraction of nucleons which contribute to the scattering.  In
the future-pointing case, the nucleons at depths $x^-$ greater than
$b^-$, $(b^- < x^- < R^- (b_T))$ contribute to the final-state rescattering;
in the past-pointing case, the nucleons at depths $x^-$ less than
$b^-$, $(-R^- (b_T) < x^- < b^-)$ contribute.  For a uniform distribution of
nucleons, this is what gives rise to the factors
\begin{align} \label{e:densint}
  \frac{1}{T(b_T)} \int\limits_{b^-}^{R^- (b_T)} dx^- \, 
\rho_N (x^-, b_T) = \frac{R^- (b_T) - b^-}{2R^- (b_T)} \notag \\
  \frac{1}{T(b_T)} \int\limits_{-R^- (b_T)}^{b^-} dx^- \, \rho_N
  (x^-, b_T) = \frac{b^- + R^- (b_T)}{2R^- (b_T)}
\end{align}
describing the fraction of nucleons which participate in the
scattering.  Here $T(b_T)$ is the number density of nucleons per unit
transverse area at impact parameter $b_T$.  Because $f_1^A$ is
$PT$-even and process-independent, it should give the same result
whether evaluated with the future-pointing gauge link of \eqref{e:GGM}
or the past-pointing gauge link of \eqref{e:GGM2}.  This is clear for
the trivial $f_1^N$ channel because both $f_1^A$ and $f_1^N$ are
invariant under the $PT$ transformation, and the longitudinal $b^-$
integral is also preserved
\begin{align} 
  \label{e:f1-2}
  \int d b^- \, e^{-\frac{1}{4} r_T^2 Q_s^2 \left(\frac{R^- (b_T) -
        b^-}{2 R^- (b_T)} \right) \ln\tfrac{1}{r_T\Lambda} } & \, W_{unp}
  \left[ (P^+ b^-)^2 \right] = \Big( b^- \rightarrow (- b^-) \Big)
 \notag \\ &=
 \int d b^- \, e^{-\frac{1}{4} r_T^2 Q_s^2 \left(\frac{b^- + R^- (b_T)}{2
       R^- (b_T)} \right) \ln\tfrac{1}{r_T\Lambda}} \, W_{unp} \left[ (P^+ b^-)^2 \right]
\end{align}
by a simple change of variables.  The Sivers function $f_{1T}^{\bot
  N}$ , on the other hand, is $PT$-odd and changes sign between the
two processes, but its associated longitudinal $b^-$ integral
\begin{align} 
  \label{e:f1-3}
  \int d b^- \, & e^{-\frac{1}{4} r_T^2 Q_s^2 \left(\frac{R^- (b_T)
        - b^-}{2 R^- (b_T)} \right) \ln\tfrac{1}{r_T\Lambda} } \, (P^+
  b^-) \, W_{OAM} \left[ (P^+ b^-)^2 \right] = \Big( b^- \rightarrow
  (- b^-) \Big)
 \notag \\ &=
 - \int d b^- \, e^{-\frac{1}{4} r_T^2 Q_s^2 \left(\frac{b^- + R^-
       (b_T)}{2 R^- (b_T)} \right) \ln\tfrac{1}{r_T\Lambda} } \, (P^+
 b^-) \, W_{OAM} \left[ (P^+ b^-)^2 \right]
\end{align}
also changes sign.  Thus both channels of \eqref{e:f1-1} are overall
even under $PT$, yielding the process-independent nuclear TMD $f_1^A$
as required.  Similar to \cite{Kovchegov:2013cva}, the
$(\vec{L}\cdot\vec{S}_N)$ spin-orbit channel here requires at least
one additional rescattering to be nonzero; in the limit $r_T^2 Q_s^2
\ll 1$, corresponding to either large transverse momentum or low
densities, the multiple scattering factor $S(r_T, b_T)$ can be
neglected, and the $b^-$ integral \eqref{e:f1-3} vanishes.  The depth
dependence on $b^-$ of the spin-orbit factor $(P^+ b^-) (\ul{p} \times
\ul{S})$, coupled to the angular dependence of the Sivers function,
provides a second $PT$-odd factor $(P^+ b^-) (\ul{p} \cdot \ul{k}')$
which enables a $PT$-odd nucleonic TMD like $f_{1T}^{\bot N}$ to
contribute to a $PT$-even nuclear TMD like $f_1^A$.  As we will show,
the same $(\vec{L}\cdot\vec{S}_N)$ spin-orbit coupling in the nucleus
also generates contributions to the $PT$-odd nuclear Boer-Mulders
function from various $T$-even nucleonic TMDs.

To better understand our result, let us evaluate \eq{e:f1-1} for
specific models of the Wigner distribution. First let us consider the
case when the internal motion of the nucleons in the nucleus is
negligible, that is the nucleons are static with respect to the
nuclear center-of-mass (the ``standard'' GGM/MV model in saturation
physics). The corresponding unpolarized Wigner distribution is
(cf. Eq.~(65) in \cite{Kovchegov:2013cva})
\begin{align}
  \label{Wcl_stat}
  W_{unp} (\vec{p}, \vec{b}) &\rfeq \frac{3\pi^2}{R^3} \theta(R^2 -
  \vec{b}^2) \delta^3(\vec{p})
  \notag \\
  W_{unp} (p,b) &= \frac{g^{+-} \, (2 \, \pi)^3}{2 \, A} \, \rho ({\ul b},
  b^-) \, \delta^2 (\ul{p}) \, \delta \left( p^+ - \frac{P^+}{A}
  \right)
\end{align}
with the constant nucleon number density (per $d^2 b_\perp \, d b^-$)
\begin{align}
  \label{eq:dens}
  \rho \left( \ul{b} , b^-\right) = \frac{\theta (R^- (\ul{b}) -
    |b^-|)}{2 R^- (\ul{b})} \, T(\ul{b}).
\end{align}
(The normalization for $W_{unp}$ is defined in \eq{Wunp_norm}.)
Since the nucleons are static, $W_{OAM} =0$ and the second term on the
right-hand side of \eq{e:f1-1} vanishes. Finally we take for the
unpolarized quark TMD of the nucleon the lowest-order
perturbative expression for a quark target
\cite{Itakura:2003jp,Meissner:2007rx}
\begin{align}
  \label{eq:unpol_LO}
  f_1^N (x, k_T) = \frac{\as \, C_F}{2 \, \pi^2 \, k_T^2} \,
  \frac{1+x^2}{1-x}
\end{align}
which we for simplicity consider in the $k_T \gg m_N$ limit. As usual $C_F = (N_c^2
-1)/(2 N_c)$ is the fundamental Casimir operator of SU($N_c$). Using
\eqref{eq:unpol_LO}, \eqref{Wcl_stat} and \eqref{e:GGM} in \eq{e:f1-1}
yields
\begin{align}
  \label{eq:f1A}
  f_1^A (x, k_T) &= \frac{N_c}{\as \, 4 \pi^4} \, \frac{1+{\hat
      x}^2}{1-{\hat x}} \, \int d^{2} b \, d^2 r \, e^{-i \ul{k} \cdot
    \ul{r}} \frac{1}{r_T^2} \, \left[ 1 - \exp \left( -\frac{1}{4}
      r_T^2 Q_s^2(b_T) \, \ln \frac{1}{r_T \, \Lambda} \right) \right].
\end{align}
The low-$x$ (and low $\hat x = A x$) limit of \eq{eq:f1A} is in
complete agreement with the previous calculation of a similar quantity
in \cite{Itakura:2003jp} (see Eqs.~(22) and (23) there keeping in mind
that unintegrated quark distribution in \cite{Itakura:2003jp} is
defined per $d k_T^2 \, dx /x$ unit of phase space, while TMDs are
conventionally defined per $d^2 k_T \, dx$). The quark saturation
scale in \eq{eq:f1A} is $Q_s^2(b_T) = 4 \pi \as^2 \, T(b_T) \, C_F
/N_c$.  Note that the agreement between our formula \eqref{e:f1-1} for
the unpolarized quark distribution with the results obtained from a
different method in \cite{Itakura:2003jp} validates our approach and
suggests that other observables, like the unpolarized gluon
distribution, when calculated in our technique would agree with the
results in the literature
\cite{Jalilian-Marian:1997xn,Kovchegov:1998bi,Metz:2011wb,Kovchegov:1996ty}.

As is well-known, putting the logarithm in the exponent of \eq{eq:f1A}
to one (as a negligibly slow varying function) casts the equation in a
form where the integral over $r_T$ can be carried out
analytically. This gives
\begin{align}
  \label{eq:f1A_short}
  f_1^A (x, k_T) &= \frac{N_c}{\as \, 4 \pi^3} \, \frac{1+{\hat
      x}^2}{1-{\hat x}} \, \int d^{2} b \ \Gamma \left( 0,
    \frac{k_T^2}{Q_s^2 (b_T)} \right).
\end{align}
This approximation is valid only for $k_T$ not much larger than $Q_s$.

To get a feeling for the correction to \eq{eq:f1A} resulting from
the spin-orbit coupling term in \eq{e:f1-1} assume that
\begin{align}
\label{Wcl_OAM}
W_{OAM} (p,b) \approx \xi \, \frac{g^{+-} \, (2 \, \pi)^3}{2 \, A} \,
\rho ({\ul b}, b^-) \, \frac{e^{-p_T^2/m_N^2}}{\pi \, m_N^2}
\, \delta \left( p^+ - \frac{P^+}{A} \right),
\end{align}
where $\xi$ is a parameter responsible for the strengths of spin-orbit
coupling and we neglect the small change in $p^+$ due to the
nonrelativistic orbital motion. Just like in \cite{Kovchegov:2013cva}
we approximate the nucleon Sivers function by the quark target
expression, assuming again that $k_T \gg m_N$: \cite{Meissner:2007rx}
\begin{align}
\label{Siv_LO}
f_{1T}^{\perp N} (x, k_T) = - \frac{\as^2 \, m_N^2 \, x \,
  (1-x)}{\pi^2 \, k_T^4} \, \ln \frac{k_T^2}{m_N^2}.
\end{align}
Employing Eqs.~\eqref{Wcl_OAM}, \eqref{Siv_LO} and \eqref{e:GGM} in
\eq{e:f1-1} we get the following correction to the unpolarized quark
TMD of a large nucleus:
\begin{align}
\label{Deltaf}
\Delta f_1^A (x, k_T) = - \xi \, \frac{\hat{x}^2 \, (1-\hat{x}) \,
  N_c^2 \, m^3_N}{8 \, \pi^4 \, \as^2 \, C_F^2 \, \rho} \int d^2 b \,
\left[ \half e^{-k_T^2/Q_s^2(b_T)} - \Gamma \left( 0,
    \frac{k_T^2}{Q_s^2 (b_T)} \right) \right],
\end{align}
where $\rho = \tfrac{3 A}{4 \pi R^3}$ is the nucleon number density in
the rest frame and we have made use of the $m_N \ll Q_s$ condition and
put $\ln (1/r_T \Lambda) \sim 1$ in the exponents. Comparing
Eqs.~\eqref{Deltaf} and \eqref{eq:f1A_short} we see that the
spin-orbit contribution is parametrically different by a factor of
$\xi/\as$. Without the knowledge of the (most likely) non-perturbative
parameter $\xi$ it is impossible to say whether the factor of
$\xi/\as$ indicates suppression or enhancement. If we take $\xi \sim
\as^4$ by analogy to atomic spin-orbit coupling in quantum
electrodynamics (QED), then $\xi/\as \sim \as^3$ which is
parametrically very small. However there is no good reason to justify
such a choice in QCD. Further work is needed to fully quantify the
role of spin-orbit coupling in the unpolarized quark TMD.

%-----------------------------------------------------------------------------------

\subsubsection{The Boer-Mulders Distribution $h_1^{\bot A}$}

%-----------------------------------------------------------------------------------

Similarly, the projection $\Gamma = \gamma^5 \gamma^+ \gamma_\bot^j$
in \eqref{e:unp3} selects the Boer-Mulders distribution $h_1^{\bot A}$
in the nucleus and the functions $h_1^{\bot N} \, , \, h_{1T}^N \, ,
\, h_{1T}^{\bot N}$ of the nucleon in \eqref{e:spinorb1} (see also
\eqref{e:qTMD2}).  Using this in \eqref{e:QCFact4} gives
\begin{align} 
  \label{e:BM-1}
  \epsilon_T^{j i} \frac{k_\bot^i}{M_A} h_1^{\bot A} (x, k_T) &=
  \frac{2 A \, g_{+-}}{(2\pi)^5} \, \int d^{2+} p \, d^{2-} b \, d^2 r
  \, d^2 k' \, e^{-i (\ul{k} - \ul{k'} - \hat{x} \ul{p}) \cdot \ul{r}}
  \, S_{(r_T, b_T)}^{[\infty^-, b^-]}
   \notag \\ & \times \:\:
   \bigg[ \epsilon_T^{j i} \frac{k_\bot^{\prime i}}{m_N} W_{unp} (p,b) \,
   h_1^{\bot N} (\hat{x}, k_T^\prime) + \tfrac{\gpm}{M_A} (P^+ b^-)
   p_\bot^i \epsilon_T^{i \ell} \, W_{OAM} (p,b)
   \notag \\ & \hspace{0.5cm} \times
   \bigg( \delta^{j \ell} h_{1T}^N (\hat{x}, k_T^\prime) +
   \frac{k_\bot^{\prime j} k_\bot^{\prime \ell}}{m_N^2} h_{1T}^{\bot
     N} (\hat{x}, k_T^\prime) \bigg) \bigg] \,.
\end{align}
It is convenient to separate the diagonal part $\delta^{j \ell}$ of
the quantity in parentheses,
\begin{align}
  \delta^{j \ell} h_{1T}^N (\hat{x}, k_T^\prime) +
  \frac{k_\bot^{\prime j} k_\bot^{\prime \ell}}{m_N^2} h_{1T}^{\bot N}
  (\hat{x}, k_T^\prime) &= \delta^{j \ell} \, h_1^N (\hat{x},
  k_T^\prime) + \frac{k_T^{\prime 2}}{m_N^2} \bigg(
  \frac{k_\bot^{\prime j} k_\bot^{\prime \ell}}{k_T^{\prime 2}} -
  \half \delta^{j \ell} \bigg) \, h_{1T}^{\bot N} (\hat{x},
  k_T^\prime) ,
\end{align}
where
\begin{align}
  h_1^N (\hat{x}, k_T^\prime) \equiv h_{1T}^N (\hat{x} , k_T^\prime) +
  \frac{k_T^{\prime 2}}{2 m_N^2} h_{1T}^{\bot N} (\hat{x}, k_T^\prime)
\end{align}
is the transversity TMD and $h_{1T}^{\bot N}$ is the pretzelosity.
Then
\begin{align} 
  \label{e:BM-2}
  \epsilon_T^{j i} \frac{k_\bot^i}{M_A} h_1^{\bot A} & (x, k_T) =
  \frac{2 A \, g_{+-}}{(2\pi)^5} \, \int d^{2+} p \, d^{2-} b \, d^2 r
  \, d^2 k' \, e^{-i (\ul{k} - \ul{k'} - \hat{x} \ul{p}) \cdot \ul{r}}
  \, S_{(r_T, b_T)}^{[\infty^-, b^-]}
   \notag \\ & \times \:\:
   \bigg( \epsilon_T^{j i} \frac{k_\bot^{\prime i}}{m_N} \: \Big[
   W_{unp} (p,b) \, h_1^{\bot N} (\hat{x}, k_T^\prime) \Big] +
   \tfrac{\gpm}{M_A} (P^+ b^-) p_\bot^i \epsilon_T^{i j} \:\: \Big[
   W_{OAM} (p,b) \, h_{1}^N (\hat{x}, k_T^\prime) \Big]
   \notag \\ & \hspace{0.75cm} + 
   \tfrac{\gpm}{M_A} (P^+ b^-) \frac{k_T^{\prime 2}}{m_N^2} \bigg(
   \frac{(\ul{p} \times \ul{k}' ) \, k_\bot^{\prime j} }{k_T^{\prime
       2}} - \half p_\bot^i \epsilon_T^{i j} \bigg) \Big[ W_{OAM}
   (p,b) \, h_{1T}^{\bot N} (\hat{x}, k_T^\prime) \Big] \bigg) \,.
\end{align}
Then we can contract the free index $j$ on both sides with $- k_\bot^n
\epsilon_T^{n j}$ and solve for the nuclear Boer-Mulders function to
obtain
\begin{align} 
  \label{e:BM-3}
  h_1^{\bot A} &(x, k_T) = \frac{2 A \, g_{+-}}{(2\pi)^5} \,
  \frac{M_A}{k_T^2} \, \int d^{2+} p \, d^{2-} b \, d^2 r \, d^2 k' \,
  e^{-i (\ul{k} - \ul{k'} - \hat{x} \ul{p}) \cdot \ul{r}} \, S_{(r_T,
    b_T)}^{[\infty^-, b^-]}
   \notag \\ & \times \:\:
   \bigg( \frac{ (\ul{k} \cdot \ul{k}')}{m_N} \: \Big[ W_{unp} (p,b)
   \, h_1^{\bot N} (\hat{x}, k_T^\prime) \Big] - \tfrac{\gpm}{M_A}
   (P^+ b^-) (\ul{p} \cdot \ul{k}) \:\: \Big[ W_{OAM} (p,b) \, h_{1}^N
   (\hat{x}, k_T^\prime) \Big]
   \notag \\ & \hspace{0.5cm} -
   \tfrac{\gpm}{M_A} (P^+ b^-) \frac{k_T^{\prime 2}}{m_N^2} \bigg(
   \frac{(\ul{p} \times \ul{k}' ) \, (\ul{k} \times \ul{k}')
   }{k_T^{\prime 2}} - \half (\ul{p} \cdot \ul{k}) \bigg) \Big[
   W_{OAM} (p,b) \, h_{1T}^{\bot N} (\hat{x}, k_T^\prime) \Big] \bigg)
   \,.
\end{align}

Thus we see that the same $\vec{L}\cdot\vec{S}_N$ coupling which
allows the quark Sivers function $f_{1T}^{\bot N}$ of the nucleons to
mix into the unpolarized quark distribution $f_1^A$ of the nucleus in
\eqref{e:f1-1} also allows the transversity $h_1^N$ and pretzelosity
$h_{1 T}^{\bot N}$ of the nucleons to mix into the nuclear
Boer-Mulders function $h_1^{\bot A}$ of the nucleus in \eqref{e:BM-3}.
The Boer-Mulders function, like the Sivers function, is a $PT$-odd
function that changes sign between the future-pointing gauge link
\eqref{e:GGM} and the past-pointing gauge link \eqref{e:GGM2}.  The
trivial channel which builds $h_1^{\bot A}$ from $h_1^{\bot N}$
already has the necessary properties under $PT$, and the longitudinal
$b^-$ integral is the same as in \eqref{e:f1-2}.  But there are also
$PT$-even channels - the nucleonic transversity $h_1^N$ and
pretzelosity $h_{1T}^{\bot N}$ - which mix into the $PT$-odd nuclear
Boer-Mulders function through the role of orbital angular momentum.
Just like in the case of the nucleonic Sivers function $f_{1T}^{\bot
  A}$ mixing into the unpolarized quark distribution $f_1^A$ of the
nucleus, the necessary reversal of $PT$ symmetry is provided by the
depth dependence $(P^+ b^-)$ in the longitudinal integral
\eqref{e:f1-3}.  And just like before, this mixing requires at least
one rescattering to provide this extra $PT$-odd factor, so it vanishes
in the limit of large $k_T$ or low charge densities.

%|||||||||||||||||||||||||||||||||||||||||||||||||||||||||||||||||||||||||||||||||||

\section{Evolution}
\label{Evolution}

Our goal in this Section is to include the quantum evolution
corrections to the quasi-classical TMDs like those calculated
above. We will separately consider the cases of $x = {\cal O} (1)$
(large-$x$) and $x \ll 1$ (small-$x$).

%%%%%%%%%%%%%%%%%%%%%%%%%%%%%%%%%%%%%%%%%%%%%%%%%%%%%%%%%%%%%%%%%%%%%%

\subsection{Large-$x$ Evolution}
\label{large-x}

The TMDs calculated above and in \cite{Kovchegov:2013cva} were found
for ${x} \sim {\cal O} (1)$: here we need to find the quantum
evolution corrections to a generic TMD given by \eq{e:QCFact3}. Note
that we are working in the $s \sim Q^2 \gg k_\perp^2$ regime, where
$s$ is the center-of-mass energy squared per nucleon; hence our ${x}$
in this section (and above) is order-one, but does not closely
approach unity. Let us begin by concentrating on summation of leading
logarithms of energy, that is on powers of $\as \, \ln (s/k_T^2)$. To
do this we can use the established formalism of saturation physics
\cite{Mueller:1994rr,Mueller:1994jq,Mueller:1995gb,Balitsky:1996ub,Balitsky:1998ya,Kovchegov:1999yj,Kovchegov:1999ua,Jalilian-Marian:1997dw,Jalilian-Marian:1997gr,Iancu:2001ad,Iancu:2000hn}.

We work in the frame where the incoming nucleus has a large $P^+$
momentum, $P^\mu = (P^+, \half\gPM\frac{M_A^2}{P^+}, \ul{0})$, while
the incoming virtual photon has a large positive $q^-$ along with a
comparable but negative $q^+$ momentum, $q^\mu =
(-\half\gPM\frac{Q^2}{q^-}, q^-, \ul{0})$. (For definitiveness we
consider the SIDIS process, though all our discussion and conclusions
apply to DY as well.) We will work in the $A^-=0$ light-cone gauge.

%%%%%%%%%%%%%%%%%%%%%%%%%%%%%%%%%%%%%%%%%%%%%%%%%%%%%%%%%%%%%%%%%%
\begin{figure}[htb]
\centering
\includegraphics[width= 0.7 \textwidth]{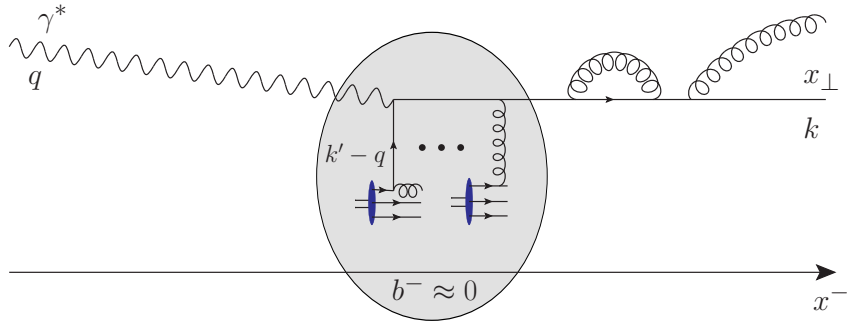}
\caption{An example of quantum evolution corrections to the SIDIS
  process and for corresponding quark TMDs.}
\label{SIDIS_evol}
\end{figure}
%%%%%%%%%%%%%%%%%%%%%%%%%%%%%%%%%%%%%%%%%%%%%%%%%%%%%%%%%%%%%%%%%%

The typical quantum evolution corrections to a SIDIS amplitude are
shown in \fig{SIDIS_evol}. Our choice of $A^-=0$ gauge prevents gluons
from being emitted or absorbed by the partons in the incoming nucleons
or by the struck quark (carrying momentum $k'-q$ in \fig{SIDIS_evol});
all these particles move predominantly in the $P^+$-direction and do
not emit gluons in the $A^-=0$ gauge in the leading-logarithmic
approximation. (We have also checked by an explicit calculation that
the emissions off of the $k'-q$ quark do not generate logarithms of
energy.) Therefore, gluon emission and absorption is limited to the
outgoing quark. Additionally, logarithms of energy can be generated
(in $A^-=0$ gauge) only by emissions which happen over long
periods of time; therefore, gluon emissions inside the nucleus do not
lead to logarithms of energy. We are left only with the gluon
emissions and absorptions by the outgoing quark after it exits the
nucleus, as shown in \fig{SIDIS_evol}.

\begin{figure}[htb]
\centering
\includegraphics[width= 0.8 \textwidth]{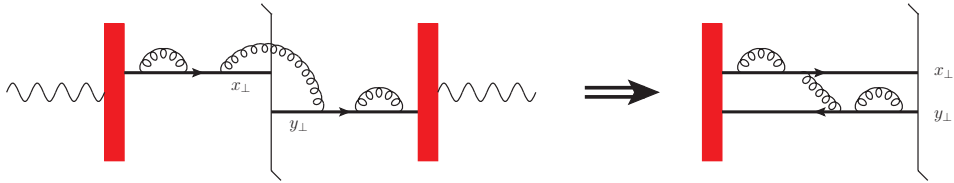}
\caption{An illustration of the crossing symmetry for the quantum
  evolution corrections on a pair of semi-infinite Wilson lines. The
  thick vertical band indicates the nucleus (the shock wave) with all
  the multiple scatterings on its nucleons.}
\label{evolution}
\end{figure}

Such emissions are easy to sum up. Using crossing symmetry, the
correction to a light-cone Wilson line at $x_\perp$ in the amplitude
and another light-cone Wilson line at $y_\perp$ in the complex
conjugate amplitude, as appears in the cross section, can be accounted
for by calculating corrections to a pair of Wilson lines in the
amplitude: one at $x_\perp$ another one at $y_\perp$. This is
illustrated in \fig{evolution} and is the basis for expressions like
\eq{e:QCFact3}, though in the quasi-classical case one resums multiple
rescattering diagrams. The crossing symmetry in \fig{evolution} is
valid both for multiple rescatterings and for gluon emission
\cite{Mueller:2012bn}.

%%%%%%%%%%%%%%%%%%%%%%%%%%%%%%%%%%%%%%%%%%%%%%%%%%%%%%%%%%%%%%%%%%
%%%%%%%%%%%%%%%%%%%%%%%%%%%%%%%%%%%%%%%%%%%%%%%%%%%%%%%%%%%%%%%%%%

We conclude that all we have to do is evolve a semi-infinite dipole
stretching in the $x^-$ direction from $0$ to $+ \infty$. (In the DY
case the dipole would stretch from $-\infty$ to $0$, but the evolution
equation would be the same.) In the leading-logarithm of energy
approximation the evolution for this object is simply given by the
(half of) virtual corrections to the standard small-$x$ evolution of
an infinite dipole
\cite{Mueller:1994rr,Mueller:1994jq,Mueller:1995gb,Balitsky:1996ub,Balitsky:1998ya,Kovchegov:1999yj,Kovchegov:1999ua,Jalilian-Marian:1997dw,Jalilian-Marian:1997gr,Iancu:2001ad,Iancu:2000hn}
and reads
\begin{align}
  \label{eq:Sevol}
  \partial_Y S_{x y} [\infty^-, b^-] (Y) = - \frac{\as \, C_F}{2 \,
    \pi^2} \int d^2 z_\perp \, \frac{(\ul{x} - \ul{y})^2}{(\ul{x} -
    \ul{z})^2 \, (\ul{z} - \ul{y})^2} \, S_{x y} [\infty^-, b^-] (Y)
\end{align} 
with the initial condition, $S_{x y} [\infty^-, b^-] (Y =0)$, given by
\eq{e:GGM} above. Here $\partial_Y \equiv \partial/\partial Y$. The
rapidity variable is given by the logarithm of energy, $Y = \ln [s \,
(\ul{x} - \ul{y})^2]$, where the IR cutoff $(\ul{x} - \ul{y})^2$ is
beyond the control of the leading logarithmic approximation considered
here. All we know is that this logarithm of energy has a cutoff of the
order of $k_T$.

\eq{eq:Sevol} is easy to solve \cite{Mueller:1994rr}. Integrating over
$z_\perp$ on the right-hand side yields
\begin{align}
  \label{eq:Ssol1}
  \partial_Y S_{x y} [\infty^-, b^-] (Y) = -\frac{\as \, C_F}{\pi} \,
  \ln \frac{(\ul{x} - \ul{y})^2}{\rho^2} \, S_{x y} [\infty^-, b^-]
  (Y)
\end{align}
with $\rho$ some ultraviolet (UV) cutoff having dimensions of
distance. Since the shortest transverse distance in the problem is of
the order of $1/Q \sim 1/s$ (since we assume that $s \sim Q^2$), we
make another approximation: keeping only the leading logarithms of
$Q^2$ we replace $\rho \to 1/Q$. In addition, remembering that we have
$s \sim Q^2$ we see that
\begin{align}
  \label{eq:log_approx}
  Y = \ln [s \, (\ul{x} - \ul{y})^2] = \ln \frac{s}{Q^2} + \ln [Q^2 \,
  (\ul{x} - \ul{y})^2] \overset{s \sim Q^2}{\approx} \ln [Q^2 \,
  (\ul{x} - \ul{y})^2].
\end{align}
In the resulting double-logarithmic approximation (DLA) (in $\ln Q^2$) we
rewrite \eq{eq:Ssol1} as
\begin{align}
  \label{eq:Ssol2}
  Q^2 \, \frac{\partial}{\partial Q^2} S_{x y} [\infty^-, b^-] (Q^2) =
  -\frac{\as \, C_F}{\pi} \, \ln [ Q^2 \, (\ul{x} - \ul{y})^2] \, S_{x
    y} [\infty^-, b^-] (Q^2).
\end{align}
The solution of \eq{eq:Ssol2} reads
\begin{align}
  \label{eq:S_DLA}
  S_{x y} [\infty^-, b^-] (Q^2) = \exp \left( -
    \int\limits^{Q^2}_{Q_0^2} \frac{d \mu^2}{\mu^2} \,
    \frac{\as \, C_F}{\pi} \, \ln [\mu^2 \, (\ul{x} - \ul{y})^2]
  \right) \, S_{x y} [\infty^-, b^-] (Q_0^2)
\end{align}
with the initial conditions $S_{x y} [\infty^-, b^-] (Q_0^2)$ at $Q_0^2 =
\frac{1}{(\ul{x}-\ul{y})^2}$ given by
\eq{e:GGM}. The exponent in \eq{eq:S_DLA} is the well-known Sudakov
form-factor \cite{Sudakov:1954sw,Collins:1989gx}, previously derived
in the saturation literature in
\cite{Levin:2010zs,Mueller:2012uf,Mueller:2013wwa,Balitsky:2014wna}. For
gluon TMDs, where the Wilson lines are adjoint, one would have to
replace $C_F \to N_c$ in \eq{eq:S_DLA}. While \eq{eq:S_DLA} is written
for a fixed coupling constant $\as$, running coupling corrections to
it can be included following
\cite{Balitsky:2006wa,Kovchegov:2006vj,Gardi:2006rp}.

Note that while we started out this derivation by studying evolution
of a semi-infinite dipole with energy, we have been working at
large-$x$ where $s \sim Q^2$ and evolution with $\ln (s/k_T^2)$ is
equivalent to evolution in $\ln (Q^2/k_T^2)$, as follows from
\eq{eq:log_approx} with the transverse momentum of quarks in the TMDs
being approximately $k_T \approx 1/|\ul{x} - \ul{y}|$. The Sudakov
form-factor \eqref{eq:S_DLA} that we have obtained is a result of
$Q^2$ evolution of the TMDs at large-$x$, and is different from the
well-known Dokshitzer--Gribov--Lipatov--Altarelli--Parisi (DGLAP)
\cite{Dokshitzer:1977sg,Gribov:1972ri,Altarelli:1977zs}
$Q^2$-evolution of the parton distribution functions (PDFs). The
difference is ultimately due to TMDs and PDFs being different objects,
with $k_T$ of the TMDs integrated over to obtain PDFs: in
\cite{Balitsky:2015qba} DGLAP evolution has been re-derived for PDFs
defined as $k_T$-integrals of TMDs.

To summarize, the QCD-evolved large-$x$ quark TMDs of a large
unpolarized nucleus in the saturation picture can be obtained from the
quark--quark correlator
\begin{align} 
  \label{e:QCFact5}
  \Phi^A & (x, \ul{k}; P; Q^2) = \frac{2 A \, g_{+-}}{(2\pi)^5} \, \int
  d^{2+} p \, d^{2-} b \, d^2 r \, d^2 k' \, e^{-i (\ul{k} - \ul{k'} -
    \hat{x} \ul{p}) \cdot \ul{r}}
  \notag \\ & \times
  \bigg( W_{unp} (p,b; P) \, \phi_{unp} (\hat{x}, \ul{k}' ; p; Q_0^2) -
  \hat{W}_{pol , \, \mu} (p,b; P) \hat\phi_{pol}^\mu (\hat{x}, \ul{k}'
  ; p; Q_0^2) \bigg) \, S_{(r_T, b_T)}^{[\infty^-, b^-]} (Q^2),
\end{align}
with the semi-infinite dipole scattering amplitude given by
\eq{eq:S_DLA} in the DLA and $S_{x y} [\infty^-, b^-] (Q_0^2)$ with
$Q_0^2 = \frac{1}{(\ul{x}-\ul{y})^2}$ given by \eq{e:GGM}. For gluon
TMDs the Sudakov form-factor would only be different by the $C_F \to
N_c$ substitution.

%%%%%%%%%%%%%%%%%%%%%%%%%%%%%%%%%%%%%%%%%%%%%%%%%%%%%%%%%%%%%%%%%%%%%%

\subsection{Small-$x$ Evolution}
\label{small-x}

\subsubsection{Evolution of Unpolarized-Target TMDs}

\begin{figure}[htb]
\centering
\includegraphics[width= 0.5 \textwidth]{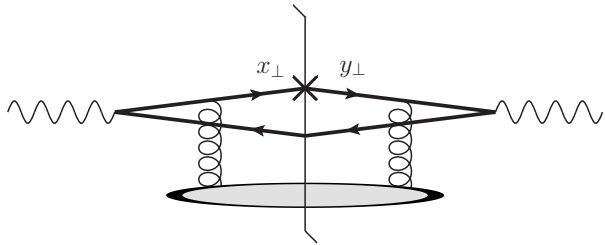}
\caption{A leading-order contribution to the unpolarized-target SIDIS process
  at small-$x$. The tagged quark is marked by a cross.}
\label{unpol_lowx_LO}
\end{figure}

Now let us consider TMD evolution in the small-$x$ regime, $s \gg Q^2
\gg k_T^2$. We begin with the unpolarized proton or nucleus TMDs,
concentrating on the unpolarized quark TMD $f_1^A$ first. As usual,
the dominant contribution at small-$x$ comes from diagrams which are
order-$\as$ suppressed compared to the channel shown in
\fig{SIDIS_qprod} above. Concentrating on the SIDIS process again, the
dominant small-$x$ contribution is pictured in
\fig{unpol_lowx_LO}. Comparing \fig{unpol_lowx_LO} to the lowest-order
(no multiple rescatterings) part of \fig{SIDIS_qprod}, one immediately
sees that the former is order-$\as$ suppressed: this is why the
channel in \fig{unpol_lowx_LO} was not considered above. However, at
small-$x$ the channel in \fig{unpol_lowx_LO} is dominant, being
enhanced by a power of $1/x$ compared to the channel in
\fig{SIDIS_qprod}.

%%%%%%%%%%%%%%%%%%%%%%%%%%%%%%%%%%%%%%%%%%%%%%%%%%%%%%%%%%%%%%%%%%
%%%%%%%%%%%%%%%%%%%%%%%%%%%%%%%%%%%%%%%%%%%%%%%%%%%%%%%%%%%%%%%%%%

One may ask why the diagram in \fig{unpol_lowx_LO} was not included
along with the evolution corrections considered in
Sec.~\ref{large-x}. Indeed the quark loop in \fig{unpol_lowx_LO}
generates a factor of $\ln (Q^2/k_T^2)$, such that the suppression of
\fig{unpol_lowx_LO} compared to the lowest-order part of
\fig{SIDIS_qprod} is diminished to a factor of $\as \, \ln
(Q^2/k_T^2)$. However it is well-known that the quark loop cannot
generate a logarithm of energy; hence the diagram in
\fig{unpol_lowx_LO} is a single-logarithmic correction, which is
beyond the DLA accuracy of the evolution in \eq{eq:S_DLA} and can be
neglected at large-$x$.

The splitting of a virtual photon into a $q \bar q$ pair in
\fig{unpol_lowx_LO} happens typically long before the interaction with
the target nucleus, denoted by a shaded oval in
\fig{unpol_lowx_LO}. Moreover, in the unpolarized SIDIS case all
interaction with the target is eikonal and, hence,
spin-independent. No interaction with one given nucleon is a special
``knockout'' interaction, where spin-dependence could be transferred
from the target to the probe, as in
\cite{Kovchegov:2013cva}. Therefore the unpolarized SIDIS cross
section at small-$x$, and the corresponding TMDs, do not have the
$b^-$ dependent effects akin to those at large-$x$ as shown in
\eq{e:QCFact3} leading to TMD mixing in Eqs.~\eqref{e:f1-1} and
\eqref{e:BM-3}. We also do not need to worry too much about the
details of nucleon dynamics encoded in the nuclear Wigner
distribution.

%%%%%%%%%%%%%%%%%%%%%%%%%%%%%%%%%%%%%%%%%%%%%%%%%%%%%%%%%%%%%%%%%%
\begin{figure}[htb]
\centering
\includegraphics[width= 0.8 \textwidth]{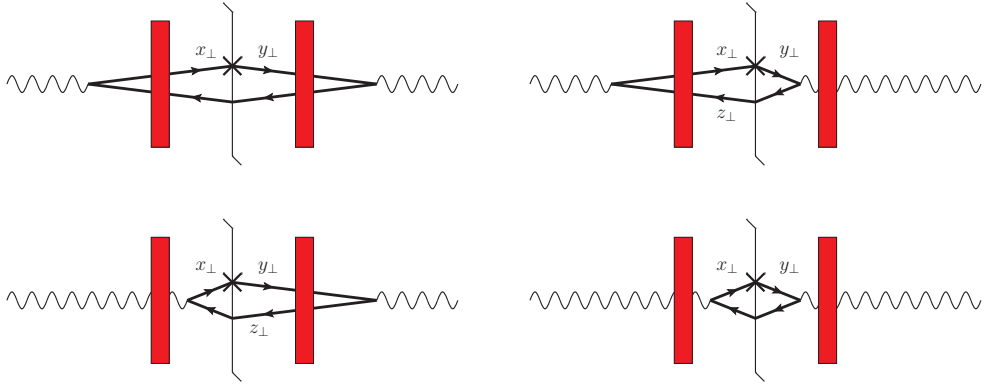}
\caption{Diagrams representing the sum of all-order contributions to
  the unpolarized-target SIDIS process at small-$x$. The tagged quark is
  marked by a cross, the thick vertical band represents the target
  nucleus (the shock wave).}
\label{unpol_lowx}
\end{figure}
%%%%%%%%%%%%%%%%%%%%%%%%%%%%%%%%%%%%%%%%%%%%%%%%%%%%%%%%%%%%%%%%%%

With this in mind, summing up the diagrams in \fig{unpol_lowx}, where
the thick vertical band represents the shock wave and we explicitly
show the four contributions where the $\gamma^* \to q \bar q$
splitting happens either before or after the photon passes through the
shock wave on either side of the cut, we use the light-cone
perturbation theory \cite{Lepage:1980fj} to write the SIDIS cross
section as (see \cite{KovchegovLevin} for a detailed presentation of a
calculation similar to this one, along with the appropriate
references)
\begin{align}
  \label{eq:qprod_smallx}
  \frac{d \sigma^{SIDIS}_{T,L}}{d^2 k_T} = \int\limits_0^1 \frac{dz}{z \, (1-z)} & \int
  \frac{d^2 x_\perp \, d^2 y_\perp \, d^2 z_\perp}{2 (2 \pi)^3} \,
  e^{-i \ul{k} \cdot (\ul{x} - \ul{y})} \, \Psi_{T,L}^{\gamma^* \to q
    \bar q} (\ul{x} - \ul{z}, z) \,
  \left[ \Psi_{T,L}^{\gamma^* \to q \bar q} (\ul{y} - \ul{z}, z) \right]^* \notag \\
  & \times \left[ S_{x,y}^{[+\infty, - \infty]} - S_{x,z}^{[+\infty, -
      \infty]} - S_{z,y}^{[+\infty, - \infty]} + 1 \right],
\end{align}
where $\Psi^{\gamma^* \to q \bar q}$ is the well-known light-cone wave
function for the $\gamma^* \to q \bar q$ splitting (normalized as in
\cite{KovchegovLevin}) given by
\begin{subequations}
\begin{align}
\label{LCwfT}
\Psi_{T}^{\gamma^* \to q \bar q} (\ul{x} , z) = & \frac{e \, Z_f}{2
  \pi} \, \sqrt{z \, (1-z)} \, \delta_{ij} \, \left[ (1 -
  \delta_{\sigma \sigma'}) \, (1 - 2 z - \sigma \, \lambda ) \, i \,
  a_f \, \frac{\ul{\epsilon}_\lambda \cdot \ul{x}}{x_\perp} \, K_1
  (x_\perp \, a_f) \right.  \notag \\ & \ \left. + \, \delta_{\sigma
    \sigma'} \, \frac{m_f}{\sqrt{2}} \,
  (1 + \sigma \lambda) \, K_0 (x_\perp \, a_f) \right], \\
\Psi_{L}^{\gamma^* \to q \bar q} (\ul{x} , z) = & \frac{e \, Z_f}{2
  \pi} \, [z \, (1-z)]^{3/2} \, \delta_{ij} \, 2 Q \, (1 -
\delta_{\sigma \sigma'}) \, K_0 (x_\perp \, a_f) \label{LCwfL}
\end{align}
\end{subequations}
for transverse ($T$) and longitudinal ($L$) polarizations of the
virtual photon. Here $z=k^-/q^-$ is the light-cone momentum fraction
of the photon carried by the tagged quark, $\sigma, \sigma', \lambda$
are the quark, anti-quark and virtual photon polarizations
respectively, $i, j$ are the quark and anti-quark colors, and $a_f^2 =
z (1-z) \, Q^2 + m_f^2$. For simplicity we assume that there is only
one quark flavor with mass $m_f$ and charge $e \, Z_f$.

$S_{x,y}^{[+\infty, - \infty]}$ in \eq{eq:qprod_smallx} is the
fundamental dipole $S$-matrix, with the light-cone Wilson lines now,
in this low-$x$ regime, stretching to both positive and negative
infinities along the $x^-$-axis. Note that the interactions of the
shock wave with the untagged anti-quark in the first diagram of
\fig{unpol_lowx} cancel on both sides of the cut.

To extract the unpolarized quark TMD we first impose the $Q^2 \gg
\perp^2$ condition on \eq{eq:qprod_smallx}, with $\perp$ denoting any
transverse momentum in the problem (this is part of the $s \gg Q^2 \gg
\perp^2$ condition defining our small-$x$ regime). It is well-known
that the large-$Q^2$ asymptotics of \eq{eq:qprod_smallx} comes from
the aligned-jet configurations dominated by either $z \to 1$ or $z \to
0$ regions of phase space
\cite{Nikolaev:1975vy,Frankfurt:1988nt}. Since we are interested in
producing a quark with $k^- \approx q^-$ we will only take the $z
\approx 1$ region into account. Note that the transverse photon
polarizations lead to a leading power of $Q^2$. The large-$Q^2$ limit
of \eq{eq:qprod_smallx} with $z \approx 1$ averaged over transverse
photon polarizations is
\begin{align}
  \label{eq:qprod_smallx3}
  & \frac{d \sigma^{SIDIS}_{T}}{d^2 k_T} = \frac{8 \, \alpha_{EM} \,
    \, Z_f^2 \, N_c}{\pi \, Q^2} \int \frac{d^2 x_\perp \, d^2 y_\perp
    \, d^2 z_\perp}{2 (2 \pi)^3} \, e^{-i \ul{k} \cdot (\ul{x} -
    \ul{y})} \, \frac{\ul{x} - \ul{z}}{|\ul{x} - \ul{z}|^2} \cdot
  \frac{\ul{y} - \ul{z}}{|\ul{y} - \ul{z}|^2} \notag \\
  & \times \frac{|\ul{x} - \ul{z}|^4 - |\ul{y} - \ul{z}|^4 - 2 \,
    |\ul{x} - \ul{z}|^2 \, |\ul{y} - \ul{z}|^2 \, \ln \frac{|\ul{x} -
      \ul{z}|^2}{|\ul{y} - \ul{z}|^2}}{(|\ul{x} - \ul{z}|^2 - |\ul{y}
    - \ul{z}|^2)^3} \left[ S_{x,y}^{[+\infty, - \infty]} -
    S_{x,z}^{[+\infty, - \infty]} - S_{z,y}^{[+\infty, - \infty]} + 1
  \right]
\end{align}
(see e.g. \cite{Kovchegov:1999kx} for details of taking the
large-$Q^2$ limit). We have put $m_f=0$ for simplicity.

%%%%%%%%%%%%%%%%%%%%%%%%%%%%%%%%%%%%%%%%%%%%%%%%%%%%%%%%%%%%%%%%%%
\begin{figure}[htb]
\centering
\includegraphics[width= 0.4 \textwidth]{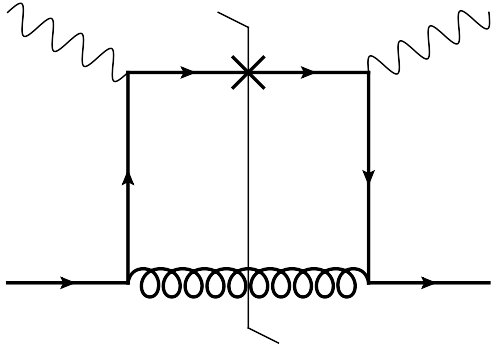}
\caption{Lowest-order diagram contributing to the SIDIS cross section
  on a single quark.}
\label{SIDIS_LO}
\end{figure}
%%%%%%%%%%%%%%%%%%%%%%%%%%%%%%%%%%%%%%%%%%%%%%%%%%%%%%%%%%%%%%%%%%

To read off the quark TMD $f_1^A$ we compare \eq{eq:qprod_smallx3} to
the lowest-order SIDIS process, modeling the target as a single
quark. The dominant contribution to this process is illustrated in
\fig{SIDIS_LO}, resulting in the lowest-order SIDIS cross section
\begin{align}
\label{eq:SIDIS_LO}
\frac{d \sigma^{SIDIS}_{T}}{d^2 k_T} = \frac{4 \pi^2 \, \alpha_{EM} \,
  Z_f^2}{s} \, f_1^{quark} (x, k_T),
\end{align}
where we have also made use of the small-$x$ limit of \eq{eq:unpol_LO}
for $f_1^{quark}$.

Comparing \eq{eq:SIDIS_LO} to \eq{eq:qprod_smallx3} and using $s
\approx Q^2/x$ we read off the small-$x$ unpolarized quark TMD for a
large nucleus to be (see \cite{Mueller:1999wm} for a similar result)
\begin{align}
\label{f1A_smallx}
& f_1^A (x, k_T) = \frac{2 \, N_c}{\pi^3 \, x} \, \int \frac{d^2
  x_\perp \, d^2 y_\perp \, d^2 z_\perp}{2 (2 \pi)^3} \, e^{-i \ul{k}
  \cdot (\ul{x} - \ul{y})} \, \frac{\ul{x} - \ul{z}}{|\ul{x} -
  \ul{z}|^2} \cdot
\frac{\ul{y} - \ul{z}}{|\ul{y} - \ul{z}|^2} \notag \\
& \times \frac{|\ul{x} - \ul{z}|^4 - |\ul{y} - \ul{z}|^4 - 2 \,
  |\ul{x} - \ul{z}|^2 \, |\ul{y} - \ul{z}|^2 \, \ln \frac{|\ul{x} -
    \ul{z}|^2}{|\ul{y} - \ul{z}|^2}}{(|\ul{x} - \ul{z}|^2 - |\ul{y} -
  \ul{z}|^2)^3} \left[ S_{x,y}^{[+\infty, - \infty]} -
  S_{x,z}^{[+\infty, - \infty]} - S_{z,y}^{[+\infty, - \infty]} + 1
\right].
\end{align}

Having the expression \eqref{f1A_smallx} for the unpolarized quark TMD
we can now determine its evolution at small-$x$. It is driven by the
small-$x$ evolution of the dipole $S$-matrix.  The small-$x$ evolution
of $S$ in the leading-logarithmic approximation (LLA) resumming powers
of $\as \, \ln (1/x)$ is well known and is given by
\cite{Balitsky:1996ub,Balitsky:1998ya,Jalilian-Marian:1997dw,Jalilian-Marian:1997gr,Iancu:2001ad,Iancu:2000hn,Weigert:2005us,Kovchegov:2008mk}
\begin{align}
  \label{eq:dip_ev}
  \partial_Y \left\langle {\hat S}_{x,y}^{[+\infty, - \infty]}
  \right\rangle_Y = \frac{\as \, N_c}{2 \pi^2} \int d^2 z_\perp \,
  \frac{(\ul{x} - \ul{y})^2}{(\ul{x} - \ul{z})^2 \, (\ul{z} -
    \ul{y})^2} \, \left[ \left\langle {\hat S}_{x,z}^{[+\infty, -
        \infty]} \, {\hat S}_{z,y}^{[+\infty, - \infty]}
    \right\rangle_Y - \left\langle {\hat S}_{x,y}^{[+\infty, -
        \infty]} \right\rangle_Y \right]
\end{align}
with $Y = \ln 1/x \approx \ln (s/Q^2)$. In \eq{eq:dip_ev} we have
separated $S_{xy} (Y) = \langle {\hat S} \rangle$ into the Wilson-line
operator ${\hat S}$ and the averaging in the nuclear wave function
evolved up to rapidity $Y$ (see \eq{e:dipop}). Unfortunately
\eq{eq:dip_ev} is not a closed integro-differential equation: the
object $\langle {\hat S} {\hat S} \rangle$ on its right-hand side is
different from $\langle {\hat S} \rangle$ on the left-hand side. The
equation closes in the large-$N_c$ limit, where it becomes
\cite{Balitsky:1996ub,Balitsky:1998ya,Kovchegov:1999yj,Kovchegov:1999ua}
\begin{align}
  \label{eq:BK}
  \partial_Y S_{x,y}^{[+\infty, - \infty]} = \frac{\as \, N_c}{2
    \pi^2} \int d^2 z_\perp \, \frac{(\ul{x} - \ul{y})^2}{(\ul{x} -
    \ul{z})^2 \, (\ul{z} - \ul{y})^2} \, \left[ S_{x,z}^{[+\infty, -
      \infty]} \, S_{z,y}^{[+\infty, - \infty]} - S_{x,y}^{[+\infty, -
      \infty]} \right],
\end{align}
where we again dropped the angle brackets. The initial condition for
\eq{eq:BK} is given by (cf. \eq{e:GGM})
\begin{align} 
  \label{e:GGM_inf}
  S_{x y}^{[+\infty, -\infty]} = \exp\left[-\frac{1}{4} |x-y|_T^2 \,
    Q_s^2 \, \ln\frac{1}{|x-y|_T \Lambda} \right].
\end{align}
Eqs.~\eqref{f1A_smallx}, \eqref{eq:dip_ev} and \eqref{eq:BK} give us
the expression for and the small-$x$ evolution of the unpolarized
quark TMD $f_1^A$.

%%%%%%%%%%%%%%%%%%%%%%%%%%%%%%%%%%%%%%%%%%%%%%%%%%%%%%%%%%%%%%%%%%
\begin{figure}[htb]
\centering
\includegraphics[width= 0.5 \textwidth]{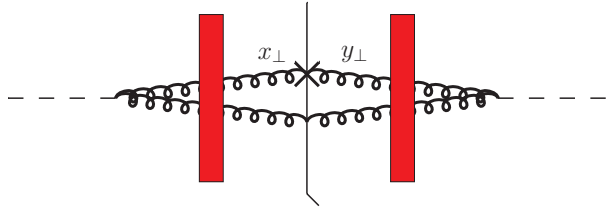}
\caption{An example of a diagram contributing a small-$x$ evolution
  correction to the unpolarized gluon TMD.}
\label{unpol_lowx_G}
\end{figure}
%%%%%%%%%%%%%%%%%%%%%%%%%%%%%%%%%%%%%%%%%%%%%%%%%%%%%%%%%%%%%%%%%%

The small-$x$ evolution of unpolarized gluon TMD is constructed in a
similar manner, with some important differences. Instead of DIS with a
photon, which only couples to quarks, consider ``DIS'' with a scalar
current $j = - (1/4) F_{\mu\nu}^a \, F^{a \, \mu\nu}$
\cite{Mueller:1989st} which couples to gluons. A gluon dipole diagram
contributing to the evolution of the unpolarized gluon TMD is shown in
\fig{unpol_lowx_G}. Note that the gluon loop brings in a logarithm of
$1/x$: hence the diagram is only a small-$x$ evolution correction to
the gluon analog of the lowest-order graph in \fig{SIDIS_LO}. To
include the small-$x$ evolution into gluon TMDs, one has to start with
such a lowest-order diagram and first include the Wilson line staple,
obtaining the Weizs\"{a}cker-Williams (WW) gluon distribution in the
classical approximation
\cite{Jalilian-Marian:1997xn,Kovchegov:1998bi,Kovchegov:2001sc,Kharzeev:2003wz,Dominguez:2010xd}. Rewriting
the semi-infinite adjoint Wilson lines in terms of (derivatives of)
the infinite Wilson lines, one can apply the JIMWLK evolution
obtaining the resulting evolution equation governing the small-$x$
asymptotics of the WW distribution, as was done in
\cite{Dominguez:2011gc,Dominguez:2011br}.

One may worry whether the diagram in \fig{unpol_lowx_G} (and other
such corrections) needs to be included at large-$x$, since the gluon
loop may give both logarithms of $Q^2$ and $s$. The resolution of this
question is in the fact that the energy logarithm coming from the
gluon loop in \fig{unpol_lowx_G} is $\ln (s/Q^2)$, which is not large
in the $s \sim Q^2$ regime of Sec.~\ref{large-x}. Hence the gluon
(\fig{unpol_lowx_G}) and quark (\fig{unpol_lowx}) loop corrections at
large-$x$ are suppressed by a power of $\as$ not enhanced by a large
logarithm, and thus are outside of the precision of our approximation.

The construction of the small-$x$ evolution of other TMDs of
unpolarized proton or nucleus, such as the quark or gluon Boer-Mulders
function, can be done similarly to the above, and are also governed by
the nonlinear evolution of the correlators of fundamental and/or
adjoint Wilson lines.

%%%%%%%%%%%%%%%%%%%%%%%%%%%%%%%%%%%%%%%%%%%%%%%%%%%%%%%%%%%%%%%%%%%%

\subsubsection{Evolution of Polarized Target TMDs: an Outline}

The evolution of polarized target TMDs at small-$x$ is somewhat
different from the unpolarized case. Here we will only give the
general outline of this evolution, with a more detailed description
left for future work.

\begin{figure}[htb]
\centering
\includegraphics[width= 0.5 \textwidth]{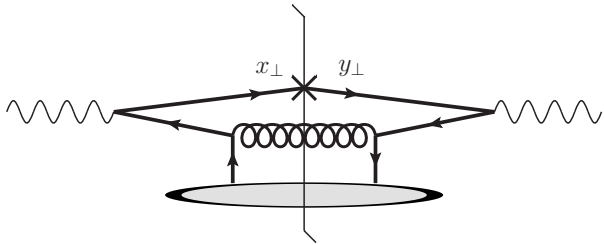}
\caption{An example of a leading-order contribution to the polarized-target
  SIDIS process at small-$x$.}
\label{unpol_lowx_LO_pol}
\end{figure}

Let us concentrate on the quark TMDs. Again the diagrams of the type
shown in \fig{unpol_lowx_LO} need to be considered, except now the
quark-gluon vertices are not eikonal, allowing for spin-dependence to
be transferred from the target nucleus to the produced quark. Note
that, as a consequence, the contribution of the diagram has a factor
of $x$ suppression compared to its eikonal contribution to the
unpolarized TMDs. These diagrams are supplemented by the same-order
(both in $\as$ and in $x$) graphs like that shown in
\fig{unpol_lowx_LO_pol}.

%%%%%%%%%%%%%%%%%%%%%%%%%%%%%%%%%%%%%%%%%%%%%%%%%%%%%%%%%%%%%%%%%%
%%%%%%%%%%%%%%%%%%%%%%%%%%%%%%%%%%%%%%%%%%%%%%%%%%%%%%%%%%%%%%%%%%

\begin{figure}[htb]
\centering
\includegraphics[width= 0.9 \textwidth]{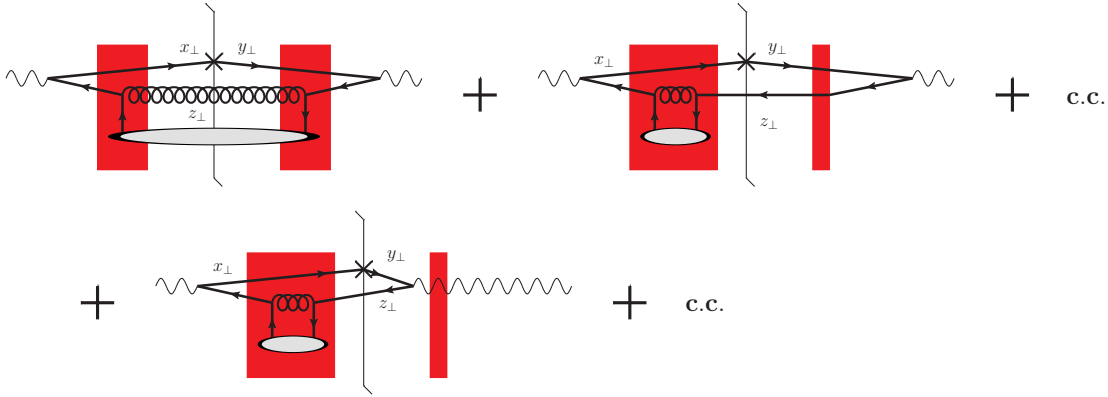}
\caption{Diagrams needed for the calculation of the polarized SIDIS
  cross section and the corresponding TMDs.}
\label{pol_mult}
\end{figure}

To include the small-$x$ evolution correction into a generic
polarized-target quark TMD one has to start by including the effects
of GGM/MV multiple rescatterings. The relevant diagrams are shown in
\fig{pol_mult}. There the red bar represents the shock wave
again. Note that we are treating the rescattering on a nucleon
carrying the spin information about the target separately, and draw it
explicitly in the graphs. (The spin-dependent scattering may also
contain gluon exchanges in the $t$-channel.) Placing such
spin-dependent rescattering inside the shock wave rectangle, as shown
in \fig{pol_mult}, implies that multiple rescatterings may happen both
before and after the spin-dependent scattering.

%%%%%%%%%%%%%%%%%%%%%%%%%%%%%%%%%%%%%%%%%%%%%%%%%%%%%%%%%%%%%%%%%%
%%%%%%%%%%%%%%%%%%%%%%%%%%%%%%%%%%%%%%%%%%%%%%%%%%%%%%%%%%%%%%%%%%

The spin-dependent scattering preserves the $z_\perp$ coordinate of
the anti-quark Wilson line. Hence all the scatterings of anti-quark
and gluon lines at $z_\perp$ cancel between the diagrams in the first
line of \fig{pol_mult}. This is similar to the reason there is no
$z_\perp$-dependence in the $S_{x,y}$ representing the interaction
with the target in the first graph of \fig{unpol_lowx}, as shown in
\eq{eq:qprod_smallx}. However, in this case we are interested in the
target polarization-dependent TMD, and cancellation of the
spin-dependent scattering implies that the TMD does not get any
contribution from the top line in \fig{pol_mult}. We are left only
with the two diagrams in the second line of \fig{pol_mult}. (Note that
the photon does not interact with the shock wave it crosses: the shock
wave to the right of the cut in the first diagram of the second line
of \fig{pol_mult} only indicates that the $\gamma^* \to q \bar q$
splitting occurs at positive light-cone time.)

\begin{figure}[htb]
\centering
\includegraphics[width= 0.9 \textwidth]{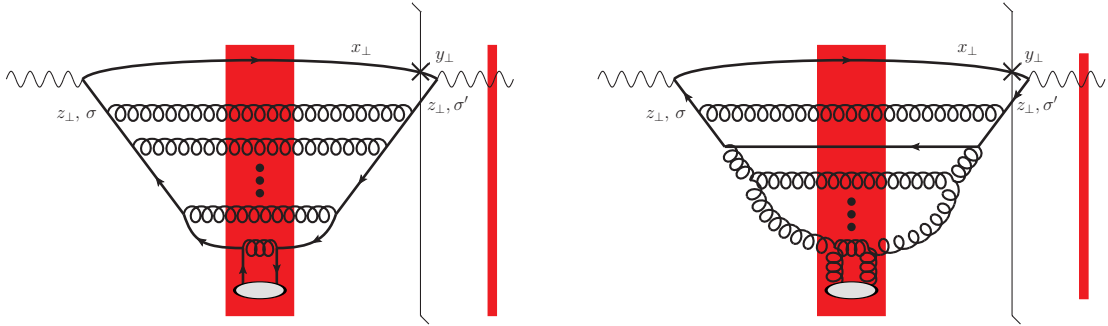}
\caption{Evolution corrections to the polarized-target SIDIS process
  at small-$x$. The gluon ladder in the right panel is of the non-BFKL
  type \cite{Bartels:1995iu,Bartels:1996wc}. The red bands represent
  the shock wave.}
\label{lowx_pol}
\end{figure}

The same arguments apply to small-$x$ evolution corrections to the
diagram in \fig{unpol_lowx_LO_pol} for the polarized TMD at hand. In
the end one is left with the diagrams where all the evolution and
interaction with the target happen entirely either to the left or to
the right of the cut, as illustrated in \fig{lowx_pol}. For simplicity
we only show the linear evolution, realized through the quark ladder
exchange shown in the left panel of \fig{lowx_pol}, and the
non-eikonal (non-BFKL) gluon ladder exchange shown in the right
panel. The nonlinear evolution corrections are represented by
interaction with the shock wave.

%%%%%%%%%%%%%%%%%%%%%%%%%%%%%%%%%%%%%%%%%%%%%%%%%%%%%%%%%%%%%%%%%%
%%%%%%%%%%%%%%%%%%%%%%%%%%%%%%%%%%%%%%%%%%%%%%%%%%%%%%%%%%%%%%%%%%

In the left panel of \fig{lowx_pol} the anti-quark in the amplitude
emits hard gluons, while itself cascading down to lower and lower $x$
until it interacts with the target. After the interaction with the
target all the emitted gluons recombine back with the anti-quark. The
same applies for the gluon ladder depicted in the right panel of
\fig{lowx_pol}. Combining each diagram with its complex conjugate we
write a general polarized-target quark TMD as
\begin{align}
  \label{eq:f1A_smallx_pol}
  f_A (x, k_T; \lambda, \Sigma) = - \int \frac{d^2 x_\perp \, d^2
    y_\perp \, d^2 z_\perp}{2 (2 \pi)^3} \, e^{-i \ul{k} \cdot (\ul{x}
    - \ul{y})} \, & \sum_{\sigma, \sigma'}
  f_{\sigma \sigma'} (\ul{x} - \ul{z}, \ul{y} - \ul{z}; \lambda) \notag \\
  & \times \, \left[ R^{\sigma \, \sigma'}_{x,z} (Y) + R^{\sigma \,
      \sigma' \, *}_{y,z} (Y) \right],
\end{align}
where $\lambda$ and $\Sigma$ are the quark and the target spin projections
correspondingly (taken in either longitudinal or transverse basis) and
$Y \approx \ln 1/x$ at small-$x$.

\eq{eq:f1A_smallx_pol} contains two main ingredients. The function
$f_{\sigma \sigma'} (\ul{x} - \ul{z}, \ul{y} - \ul{z}; \lambda)$ has
to be separately calculated for every polarized TMD (using the
$\gamma^* \to q \bar q$ light-cone wave function) by analogy to what
was done in the unpolarized target case: this is left for future
work. The interaction of the $x,z$ (and $y,z$) dipole with the target,
along with its evolution shown explicitly in \fig{lowx_pol}, is
included through the Reggeon exchange amplitude $R^{\sigma \,
  \sigma'}_{x,z} (Y)$ ($R^{\sigma \, \sigma' \, *}_{y,z} (Y)$). The
evolution of $R^{\sigma \, \sigma'}_{x,z} (Y)$ involves mixing between
the quark and gluon ladders, as shown in the right panel of
\fig{lowx_pol}. It is possible that the evolution of the (non-BFKL)
gluon ladder depends on the TMD in question.  For example, it is
conceivable that the evolution of the quark transversity could be
different from the quark helicity. While further investigation of this
evolution for various polarized TMDs is left for future work, let us
give an example of such evolution by considering only the quark
sector.

By the quark sector we mean considering only the diagram in the left
panel of \fig{lowx_pol}, without any mixing with the gluon
ladders. The quark ladder evolution is built up out of the $q \to q G$
splittings with the quark after the splitting being the softer
particle. Corresponding small-$x$ evolution kernel is diagonal in
quark helicity. Hence the large-$N_c$ nonlinear evolution of
$R^{\sigma \, \sigma'}_{x,z} (Y)$ for a ladder with quarks in the
$t$-channel is independent of the anti-quark helicities $\sigma, \,
\sigma'$ on either side of the cut, and in the large-$N_c$ limit can
be simply read off from the evolution of the flavor non-singlet
structure function in \cite{Itakura:2003jp} (see also
\cite{Kirschner:1983di,Kirschner:1994vc,Kirschner:1994rq,Griffiths:1999dj,Bartels:1995iu,Bartels:1996wc,Bartels:2003dj}):
\begin{align}
  \label{eq:Revol}
  R^{\sigma \, \sigma'}_{x,z} (Y) = r^{\sigma \, \sigma'}_{x,z} (Y) +
  \frac{\as \, C_F}{2 \pi^2} \, \int \frac{d^2 w_\perp}{w_\perp^2} \,
  \int\limits_{Y_i}^{\mbox{min} \left\{ Y, Y - \ln \left[ (\ul{z} -
        \ul{x})^2/w_\perp^2 \right] \right\}} \, dy \, R^{\sigma \,
    \sigma'}_{z,z+w} (y) \, S_{x,z}^{[+\infty, - \infty]} (y).
\end{align}
Here $Y_i$ is some initial rapidity for the evolution and $r^{\sigma
  \, \sigma'}_{x,z} (Y)$ is the initial condition. This initial
condition is given by \cite{Itakura:2003jp}
\begin{align}
  \label{eq:init}
  r^{\sigma \, \sigma'}_{x,y} (Y) = \frac{s}{2} \, \frac{d {\hat
      \sigma^{\sigma \, \sigma'}}(Y)}{d^2 b} \, \exp\left[-\frac{1}{4}
    |x-y|_T^2 \, Q_s^2 \, \ln\frac{1}{|x-y|_T \Lambda} \right]
\end{align}
with $d {\hat \sigma^{\sigma \, \sigma'}} (Y)/d^2 b$ the cross section
of lowest-order spin-dependent scattering of the $x,y$ dipole on one
nucleon in the nucleus (here $\ul{b} = (\ul{x} + \ul{y})/2$). This
cross section has to be calculated separately for each TMD.

To solve \eq{eq:Revol} one has to first solve \eq{eq:BK} to find
$S_{x,z}^{[+\infty, - \infty]} (y)$. Hence the equation
\eqref{eq:Revol} actually mixes the quark and (BFKL) gluon ladders,
albeit at the non-linear level.

\eq{eq:Revol} is double-logarithmic in energy, that is, it resums
powers of $\as \, \ln^2 s$. Hence it contains $R$, which resums
double-logarithms of energy, and $S$, resumming single logarithms
\cite{Itakura:2003jp}. Indeed, for consistency, the evolution for $R$
has to be augmented by the single-logarithmic term: this is left for
the future work. Eqs.~\eqref{eq:f1A_smallx_pol} and \eqref{eq:Revol}
complete our outline for the general form of the small-$x$ evolution
of the polarized quark TMDs.

Note again that the full QCD Reggeon evolution (even at the linear
level) also includes gluon ladders, mixing with the quark ladders, as
shown in the right panel of \fig{lowx_pol}. There one of the gluons
carries the information about the polarization of the target
\cite{Bartels:1995iu,Bartels:1996wc}. Hence \eq{eq:Revol} is
incomplete, and has to be augmented by a mixing term with the gluon
ladder along with a separate evolution for the gluon ladder (also
mixing with the quark ladder), similar to how it was done in
\cite{Bartels:1995iu,Bartels:1996wc}. While this is outside the scope
of this work, here we note that the $x$-dependence found in
\cite{Bartels:1996wc} scales as
\begin{align}
\label{lowx-sc}
R^{\sigma \, \sigma'} \sim x^{-z_s \, \sqrt{\frac{\as \, N_c}{2 \pi}}},
\end{align}
with the numerical factor $z_s = 3.45$ for 4 quark flavors. The power
in \eq{lowx-sc} is large and negative, and can easily become large
enough to make the net power of $x$ smaller than $-1$ for the
realistic strong coupling of the order of $\as = 0.2 - 0.3$, resulting
in polarized TMDs which actually {\sl grow} with decreasing $x$ fast
enough for the integral of the TMDs over the low-$x$ region to be
(potentially) large. The tantalizing possibility of this growth
generating a significant contribution to the proton spin coming from
low-$x$ partons will be explored in the future work.

%|||||||||||||||||||||||||||||||||||||||||||||||||||||||||||||||||||||||||||||||||||

%|||||||||||||||||||||||||||||||||||||||||||||||||||||||||||||||||||||||||||||||||||

\section{Conclusions}

%|||||||||||||||||||||||||||||||||||||||||||||||||||||||||||||||||||||||||||||||||||

In this work, we have presented calculations for the quark TMDs of an
unpolarized target in the quasi-classical approximation and with
leading-logarithmic quantum evolution.  At face value the calculation
applies to a heavy nucleus under the resummation $\alpha_s^2 A^{1/3}
\sim \ord{1}$, but its core features should also be valid for a
high-energy proton with a large parton density generated by quantum
evolution.  Attempts to model the dense proton as a ``nucleus'' have
been successfully applied in the past to phenomenology in inclusive
deep inelastic scattering, so this approach may be a valuable tool in
studying the TMDs of the proton as well.  Regardless, the formalism
presented here makes it possible to perform first-principles
calculations of TMDs and spin-orbit structure within a controlled
resummation of QCD.

The primary results of our calculation are as follows.  The
quasi-classical factorization formula \eqref{e:QCFact3} expresses the
relationship between the quark TMDs of the nucleus, the quark TMDs of
the nucleons, the distribution of nucleons in the nucleus, and the
multiple rescattering on spectator nucleons.  The parameterization
\eqref{e:Wig9} expresses the most general form of the nucleon Wigner
distribution which is consistent with rotational invariance, parity,
and time reversal.  Together, these yield the decomposition of the
unpolarized quark distribution $f_1$ \eqref{e:f1-1} and the quark
Boer-Mulders distribution $h_1^{\bot}$ \eqref{e:BM-3} in terms of the
TMDs of the nucleons.  The quasi-classical expressions can be taken as
initial conditions for the subsequent quantum evolution.  In the
large-${x}$ regime, the leading evolution \eqref{eq:S_DLA} is
double-logarithmic, resumming powers of $\alpha_s \ln^2 Q^2$ in
agreement with the Sudakov form factor and the Collins-Soper-Sterman
evolution equation.  In the small-${x}$ regime, the unpolarized quark
distribution is given by \eqref{f1A_smallx}, and its evolution is
governed by BK-JIMWLK equations \eqref{eq:dip_ev} and \eqref{eq:BK}.
The small-${x}$ evolution suggested by the polarized TMDs appears to
be more complex, involving the familiar BK-JIMWLK evolution as an
ingredient for the more intricate Reggeon evolution \eqref{eq:Revol}
(augmented by the mixing with the gluon ladders, as shown in
\fig{lowx_pol}).

Clearly further investigation of the perturbative QCD Reggeon
evolution is warranted in the future. If the Reggeon evolution leads
to polarized TMDs that may grow at small-$x$ faster than $1/x$, as
\eq{lowx-sc} with the power of $x$ taken from \cite{Bartels:1996wc}
appears to suggest, one should try to include saturation effects into
this evolution. The inclusion of saturation effects may make the
integral of the resulting TMD over small-$x$ convergent, and would
allow one to assess the size of the contribution of such an
integral. A large contribution coming from the small-$x$ region may
help resolve the proton spin puzzle by identifying the phase space
region containing the missing spin: this possibility is important and
has to be explored in the future.

In our formalism, essentially all of the model dependence is
encapsulated into the nonperturbative Wigner distribution of nucleons
inside the nucleus.  This distribution, however, is highly constrained
by symmetry, making it possible to identify distinct channels which
couple the TMDs of the nucleus to the TMDs of the nucleons.  Our
approach is also highly amenable to modeling and phenomenology.  First
one chooses an ansatz for the Wigner distribution such as
\eqref{Wcl_stat}; this then determines the form of the quasi-classical
multiple scattering factor through the integral \eqref{e:densint}.
With these ingredients, one can evaluate the intermediate integrals
over $p, b$, and $r$ which couple the nuclear TMDs to the nucleonic
ones.  If desired, one can then also choose initial conditions for the
nucleonic TMDs, such as the quark target model or scalar diquark model
of \cite{Meissner:2007rx}, allowing for explicit analytic or numerical
calculation of the nuclear TMDs.  These functional forms can then be
evolved with the appropriate quantum evolution equations to the
kinematics appropriate for comparison with experimental data.  In
future work, we would like to extend our computation of TMDs in the
dense limit to the full set of leading-twist quark and gluon TMDs.
This would provide another theoretical benchmark for understanding the
range of spin-orbit physics permitted by QCD and would be well-suited
to a global fit of available TMD data.

One novel feature of our calculation is that, due to the possibility
of spin-orbit coupling in the distribution of nucleons, there can be
mixing between different TMDs at the level of the nucleus and the
nucleons. In \eqref{e:f1-1} the nucleonic Sivers function mixed into
the nuclear quark distribution, and in \eqref{e:BM-3} the nucleonic
transversity and pretzelosity mixed into the nuclear Boer-Mulders
function.  All three of these mixings are due to the same
$\vec{L}\cdot\vec{S}_N$ spin-orbit correlation with strength given by
$W_{OAM}$.  The fact that the same underlying correlation is
responsible for multiple mixings is a testable prediction of the
theory: in principle, if the unpolarized quark distribution $f_1^N$
and Sivers function $f_{1T}^{\bot N}$ of the nucleon are known from
experiment, then a further measurement of the unpolarized quark
distribution $f_1^A$ of the nucleus can allow an extraction of the
spin-orbit coupling term $W_{OAM}$ which is present in the nucleus.
This would then provide a prediction for the amount of mixing that
should occur between the nucleonic transversity $h_1^N$ or
pretzelosity $h_{1T}^{\bot N}$ and the Boer-Mulders function
$h_1^{\bot A}$ of the nucleus.  Such an extraction would require good
coverage of the $p_T$ dependence of the nuclear TMDs, and is thus
likely to only be accessible at a future electron-ion collider.

The TMD mixing observed here couples the $PT$-even and $PT$-odd
sectors, mediated by the spin-orbit coupling mechanism and the depth
dependence of the multiple scattering.  This is a very general feature
of TMDs in the dense limit which goes beyond the specific
$\vec{L}\cdot\vec{S}_N$ coupling that can occur in an unpolarized
nucleus.  A similar mechanism $\vec{L}\cdot\vec{S}_A$ was already
observed for the case of a transversely polarized nucleus, resulting
in the mixing of the nucleonic unpolarized quark distribution into the
nuclear Sivers function \cite{Kovchegov:2013cva}.  Similar features
should again occur in the general case of a polarized nucleus, with a
relatively small number of possible spin-orbit correlations generating
a relatively large number of mixings between the nuclear and nucleonic
TMDs.  Indeed, one can imagine performing the same sort of analysis
for the ``generalized transverse-momentum-dependent parton
distribution functions'' (GTMDs) which are analogous operators to
\eqref{e:qcorr1}, taken between off-forward matrix elements of
hadronic states (see, for example, \cite{Meissner:2009ww}).  The GTMDs
are the ``mother functions'' from which one can obtain both the TMDs
and the generalized parton distributions (GPDs).  If the same type of
mixing occurs at the level of the GTMDs, it raises the interesting
possibility of a single spin-orbit coupling term being responsible for
different mixings in both the TMD and GPD sectors.  The methodology
presented here, which has at its foundation a genuine resummation of
QCD, offers a new approach to the calculation of TMDs and related
quantities, with bountiful applications to both theory and
phenomenology.

%%%%%%%%%%%%%%%%%%%%%%%%%%%%%%%%%%%%%%%%%%%%%%%%%%%%%%%%%%%%%%%%%%%%%%%%%%%%%%%
%%%%%%%%%%%%%%%%%%%%%%%%%%%%%%%%%%%%%%%%%%%%%%%%%%%%%%%%%%%%%%%%%%%%%%%%%%%%%%%

\section*{Acknowledgments}

The authors are grateful to Elke Aschenauer, Ian Balitsky, Stan
Brodsky, C\'{e}dric Lorc\'{e}, Daniel Pitonyak, Jianwei Qiu, Andrey
Tarasov, and Yi Yin for informative discussions. This material is
based upon work supported by the U.S. Department of Energy, Office of
Science, Office of Nuclear Physics under Award Number DE-SC0004286.
MS is supported under DOE Contract No. DE-SC0012704. \\

%%%%%%%%%%%%%%%%%%%%%%%%%%%%%%%%%%%%%%%%%%%%%%%%%%%%%%%%%%%%%%%%%%%%%%%%%%%%%%%%%%%%%%
%%%%%%%%%%%%%%%%%%%%%%%%%%%%%%%%%%%%%%%%%%%%%%%%%%%%%%%%%%%%%%%%%%%%%%%%%%%%%%%%%%%%%%

%###################################################################################

\appendix

\section{Rotational Invariance on the Light Front}
\label{sec:Karmanov}

In Sec.~\ref{sec:paramet}, we discussed the symmetries of the Wigner
distribution of nucleons in the nucleus, including the naive
expectation that the Wigner distribution should be rotationally
invariant in the rest frame of the nucleus.  Indeed, the Wigner
distribution is independent of the preferred axis contained in the
Wilson lines and virtual photon.  But there is a subtlety which we
will now address: since the Wigner distribution is defined in terms of
the light-front wave functions, there {\it is} a special direction
given by the {\it quantization axis} with respect to which those wave
functions are defined.  In the standard formulation of light-front
perturbation theory, the system is quantized at fixed $x^+ \! \sim \!
(c t + z)$, so that three-dimensional rotation invariance is broken by
the special role of the $z$-axis.\footnote{Note that in this Appendix
  we will use both $c=1$ and $c\neq 1$ conventions interchangeably: we
  have taken special care to not cause any confusion with this
  notation.} As discussed in \cite{Brodsky:1997de}, these light-front
wave functions are especially well-suited to describing high-energy
collision dynamics along the $z$-axis, since they are invariant under
longitudinal boosts, transverse boosts, and rotations in the
transverse plane.  But rotations which change the longitudinal
direction are ``dynamical''; that is, they invoke the interaction
Hamiltonian and can therefore change the particle content of the
system.  The special role of the $z$-axis also breaks invariance under
ordinary parity $P$ and time-reversal $T$, necessitating the
application of the modified discrete symmetries ``light-cone parity''
$P_\bot$ and ``light-cone time reversal'' $T_\bot$ which preserve the
$z$-axis and satisfy $P_\bot T_\bot = P T$ \cite{Brodsky:2006ez}.

To describe the properties of the light-front wave functions under
three-dimensional rotations without invoking the interaction
Hamiltonian, it is necessary to use the ``covariant light-front
dynamics'' of \cite{Carbonell:1998rj} in which the quantization axis
$\omega^\mu$ is kept arbitrary.  If the theory is quantized at fixed
$\omega \cdot x$, then rotations which transform not only the physical
vectors, but also the quantization axis $\omega^\mu$ correspond to
``kinematic'' transformations which do not invoke the interaction
Hamiltonian.  Consequently, a light-front wave function for spinless
particles is a Lorentz scalar when the quantization axis $\omega^\mu$
is also transformed.  In general, the spin indices of a light-front
wave function rotate according to their group representation, but they
transform more simply in terms of the covariant spin vector $S^\mu$
\eqref{e:spinRF2}.

Defining the lightlike quantization axis as 
\begin{align}
  \omega^\mu \equiv \sqrt{\tfrac{\gPM}{2}} \, \left( 1 , - \hat{n} \right)
\end{align}
for an arbitrary spacelike unit vector $\hat{n}$, we see that the
analog of the usual ``plus'' direction is
\begin{align}
  p^+ \equiv \omega \cdot p = \sqrt{\tfrac{\gPM}{2}} \left( E_p +
    \vec{p} \cdot \hat{n} \right),
\end{align}
where $\hat{n} = \hat{z}$ in the usual approach.  The Wigner
distribution can then be written as
\begin{align} \label{e:Wig6} W_{\lambda \lambda'} (\bar p , b ; P ;
  \omega) = \frac{1}{2(2\pi)^3} \int \frac{d^{2+}(\delta p)}{\sqrt{p^+
      p^{\prime +}}} \, e^{-i (\delta p) \cdot b} \, \psi_\lambda^*
  (p' ; P ; \omega) \: \psi_{\lambda'} (p ; P ; \omega),
\end{align}
where $b^+ \equiv \omega \cdot b
\equiv 0$ and the wave functions now explicitly depend on the
quantization axis $\omega^\mu$.  

Since the ``longitudinal direction'' is now defined by $\hat{n}$, the
longitudinal momentum fraction $\alpha$ of a nucleon with
momentum $p^\mu$ in a nucleus with momentum $P^\mu$ is 
\begin{align}
    \alpha \equiv \frac{\omega \cdot p}{\omega \cdot  P} = \frac{p^+}{P^+},
\end{align}
and the vector $R^\mu \equiv p^\mu - \alpha P^\mu$, defined so that $\omega
\cdot R = 0$, selects out the momentum transverse to $\hat{n}$:
\begin{align}
  R^2 = (R^0)^2 - \vec{R}^2 = - \left[ \vec{R}^2 - (\vec{R} \cdot
    \hat{n})^2 \right] \equiv - R_T^2.
\end{align}
The variables $\alpha, R^2$ are thus scalars under the generalized
transformations which include $\omega^\mu$.  As shown in
\cite{Carbonell:1998rj}, the wave functions depend only on the two
invariants: $\psi = \psi (\alpha , R^2)$.  In terms of these
quantities, we can rewrite the Wigner distribution \eqref{e:Wig6} as
\begin{align}
  W_{\lambda \lambda'} (\bar p , b ; P ; \omega) = \frac{1}{2(2\pi)^3}
  \int d^2(\delta R) \frac{d (\delta \alpha)}{\sqrt{\alpha \alpha^{\prime}}} \, e^{-i
    (\delta \alpha) P \cdot b} \, e^{-i (\delta R) \cdot b} \psi_\lambda^*
  (\alpha' , R^{\prime 2}) \: \psi_{\lambda'} (\alpha , R^2),
\end{align}
with $\delta R^\mu = R^\mu - R^{\prime \mu}$ and $\delta \alpha =
\alpha - \alpha'$, and project out the polarized and unpolarized
distributions using \eqref{e:Wig3} to obtain
\begin{align}
  W({ p}^\mu , b^\mu , S^\mu ; P^\mu ; \omega^\mu) = W_{unp}
  ({ p}^\mu , b^\mu ; P^\mu ; \omega^\mu) - S_\nu \,
  \hat{W}_{pol}^\nu ({ p}^\mu , b^\mu ; P^\mu ; \omega^\mu)
\end{align}
Thus the Wigner distribution is now manifestly invariant under the
generalized Lorentz transformations.

The rotation properties of the wave functions are seen most clearly in
the ``constituent rest frame'' (CRF).  In light-front perturbation
theory, there is no ``rest frame'' in the conventional sense, since
one component of the momentum $(p^-)$ is not conserved.  It is thus
impossible to simultaneously set the net spatial momentum $\vec{P}$ of
the nucleus and $\sum_i \vec{p}_i$ of the nucleons to zero.  Instead,
the momenta satisfy a modified conservation law,
\begin{align} \label{e:LFcons}
  \sum_i p_i^\mu = P^\mu + \tau \omega^\mu,
\end{align}
where $\tau$ is the deviation from the ``energy shell.''  In the
constituent rest frame, one sets the net spatial momentum of the
constituents to zero, $\sum_i \vec{p}_i = 0$, with the parent particle
forced to have nonzero momentum, $\vec{P} = \tau \hat{n}$.

In the constituent rest frame, the variables $\alpha, R^2$ simplify:
\begin{align}
  -R^2 &\crfeq (\vec{p} - \alpha \tau \hat{n})^2 - \left[
    (\vec{p}\cdot\hat{n}) - \alpha \tau \right]^2
  =
  \vec{p}^2 - (\vec{p}\cdot\hat{n})^2
  \notag \\
  \alpha &\equiv \frac{\omega \cdot p}{\omega \cdot P} 
    = \frac{\omega \cdot p}{\sum_i \omega \cdot p_i} 
    = \frac{E_p + \vec{p}\cdot\hat{n}}{\sum_i E_i + \left( \sum_i
        \vec{p}_i \right) \cdot \hat{n}}
    \crfeq
    \frac{E_p + \vec{p} \cdot \hat{n}}{\sum_i E_i} ,
\end{align}
where we used \eqref{e:LFcons} and $\omega^2 = 0$.  The wave functions
$\psi(\alpha, R^2)$ can therefore be expressed as $\psi (\vec{p}^2 ,
\vec{p} \cdot \hat{n})$ which are equivalent invariants but have a
simple meaning in the constituent rest frame.  The covariant
formulation thus makes explicit the dependence on the preferred
direction $\hat{n}$; this dependence is a feature of the relativistic
wave function and in general cannot be avoided.

%***********************************************************************************
\begin{figure}[bt]
 \centering
 \includegraphics[width=0.6 \textwidth]{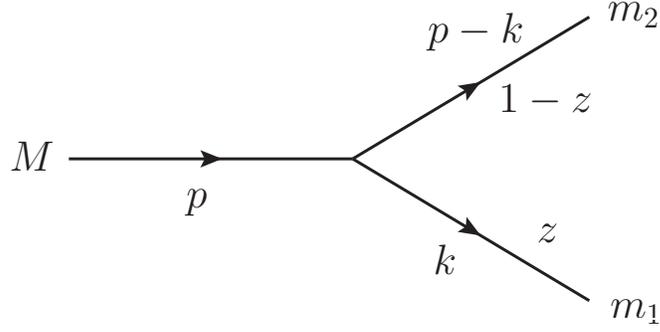}
 \caption{Simple wave function in which a scalar with mass $M$ and
   momentum $P$ splits into two scalars, one with mass $m_1$ and
   momentum $k$ and the other with mass $m_2$.}
\label{f:NRWF}
\end{figure}
%***********************************************************************************

However, there is one limit in which the dependence on the special
direction $\hat{n}$ disappears: the non-relativistic limit
$\vec{p_i}^2 \ll m^2$ or $c \rightarrow \infty$.  In this limit, the
light-front quantization condition $\omega \cdot x = c t + (\vec{x}
\cdot \hat{n}) = const.$ reduces to the equal-time condition $c t =
const.$ and the dependence on $\hat{n}$ drops out.  Consequently, in
the non-relativistic limit the light-front wave functions should no
longer depend on $\hat{n}$: $\psi (\vec{p}^2, \vec{p} \cdot \hat{n})
\rightarrow \psi(\vec{p}^2)$, restoring the rotational invariance of
the physical vectors in the rest frame.  For example, consider the
simple wave function for the splitting of massive scalar particles
shown in Fig.~\ref{f:NRWF}.  Taking the tri-scalar vertex as
$\lambda$, the standard calculation of the wave function gives
\begin{align}
  \psi(k,p) &= \frac{1}{2\gpm p^+} \: \frac{\lambda}{p^- - k^- - (p-k)^-} 
  \notag \\ &=
  \frac{- \lambda \: z (1-z)}{ (k - z p)_T^2 - z (1-z) M^2 + (1-z)
    m_1^2 + z m_2^2},
\end{align}
where $z = k^+ / p^+$ is the momentum fraction of the scalar $m_1$.
In the constituent rest frame we have
\begin{align}
  \psi( \vec{k}^2 , \vec{k} \cdot \hat{n})   &\crfeq \frac{- \lambda
    \: z (1-z)}{\vec{k}^2 - (\vec{k}\cdot\hat{n})^2 - z (1-z) M^2 +
    (1-z) m_1^2 + z m_2^2}
  \notag \\
  z &\crfeq \frac{\sqrt{\vec{k}^2 + m_1^2} + \vec{k} \cdot
  \hat{n}}{\sqrt{\vec{k}^2 + m_1^2} + \sqrt{\vec{k}^2 + m_2^2}},
\end{align}
and expanding $\vec{k}^2 \ll m_1^2 , m_2^2 , M^2$ in the
nonrelativistic limit (i.e., $(\vec{k} c)^2 \ll (m c^2)^2$ for $c
\rightarrow \infty$), we obtain
\begin{align}
  \psi(\vec{k}^2, \vec{k}\cdot\hat{n}) \approx \frac{- \lambda}{3 m^2}
  \left[ 1 - \frac{4 \vec{k}^2}{3 m^2} \right] = \psi(\vec{k}^2)
\end{align}
as desired.  Note that in the limit of nonrelativistic constituent
motion, the rest frames of the parent particle and the constituents
coincide, and we no longer need to distinguish between the
``constituent rest frame'' and the ordinary rest frame.

Indeed, the non-relativistic limit is precisely the limit which is
relevant for the orbital motion of nucleons in a heavy nucleus.  If we
expand \eqref{e:Wig6} to lowest order in the nucleon velocities $\beta
= v/c = |\vec{k}|/m_N \ll 1$, then we have
\begin{align}
  E_p &\approx m_N \left( 1 + \ord{\beta^2} \right) \notag \\
  \alpha &\approx \frac{1}{A} \left( 1 + \frac{p_\parallel}{m_N} + \ord{\beta^2} \right) \notag \\
  p_\parallel & \equiv (\vec{p} \cdot \hat{n}) \approx \Big(A \, m_N \Big) \left(\alpha - \frac{1}{A} \right) \left( 1 + \ord{\beta} \right) \notag \\
  \frac{d^{2+} (\delta p)}{\sqrt{p^+ p^{\prime +}}} &\approx \frac{d^3
    (\delta p)}{m_N} \left( 1 + \ord{\beta} \right).
\end{align}
The Fourier factor
$\exp[-i (p - p') \cdot b]$ then becomes
\begin{align}
  \exp[-i (p-p') \cdot b] &= \exp[ -i \gpm (\alpha - \alpha') P^+ b^-] \, \exp[
  +i (\ul{p} - \ul{p'}) \cdot \ul{b}]
  \notag \\ & \!\!\!\!\!\!\!\!\!\!\!\! \approx
  \exp\left[-i (p_\parallel - p^\prime_\parallel) \left( \tfrac{1}{A
        \, m_N} \gpm P^+ b^- \right) \right] \, \exp[ +i (\ul{p} -
  \ul{p'}) \cdot \ul{b}] \, \left( 1 + \ord{\beta} \right).
\end{align}
Since $b^+ \equiv 0$ and $b_\parallel = \sqrt{\frac{\gpm}{2}} \left(
  b^+ - b^- \right)$, we have
\begin{align}
  b^- &= - \sqrt{2\gPM} \, b_\parallel ,
\end{align}
and in the rest frame
\begin{align}
  P^+ \rfeq \sqrt{\frac{\gPM}{2}} \, (A \, m_N) ,
\end{align}
such that
\begin{align}
   \frac{1}{A \, m_N} \gpm P^+ b^- \rfeq - b_\parallel
\end{align}
and the Fourier factor becomes
\begin{align}
  \exp[-i (p-p') \cdot b] &\rfeq \exp[ + i (\vec{p} - \vec{p'}) \cdot
  \vec{b} ] \, (1 + \ord{\beta}).
\end{align}
Substituting these expressions back into \eqref{e:Wig6} and neglecting
the dependence of the light-front wave functions on $\hat{n}$ for
nucleons moving non-relativistically in the heavy nucleus, we obtain
\begin{align} 
  \label{e:Wig7}
  W_{\lambda \lambda'} (\vec{\bar p} , \vec{b}) \rfeq \frac{1}{2
    (2\pi)^3} \int \frac{d^3 (p - p')}{m_N} \, e^{+ i (\vec{p} -
    \vec{p'}) \cdot \vec{b}} \: \psi_\lambda^* \big( {\vec{p'}}^2
  \big) \, \psi_{\lambda'} \big( {\vec p \,}^2 \big).
\end{align}
Once again projecting out the polarized and unpolarized structures using
\eqref{e:Wig3}, we obtain
\begin{align} 
  \label{e:Wig8}
  W(\vec{ p} , \vec{b} , \vec{S}) \rfeq W_{unp} (\vec{p} ,
  \vec{b}) + \vec{S} \cdot \vec{W}_{pol} (\vec{p}, \vec{b}),
\end{align}
which recovers the naive expectation of manifest three-dimensional
rotation invariance in the nuclear rest frame.

The key is the change of variables between the boost-invariant
quantities like $\bar{\alpha}, P^+ b^-$ and the 3-vectors $\vec{ \bar
  p} , \vec{b}$:
\begin{align} 
  \label{e:NRvars}
  \vec{ \bar p} &\rfeq \left[ \, \ul{ \bar p} \, , \, M_A \! \left(
      \bar{\alpha} - \frac{1}{A} \right) \, \right]
  \notag \\
  \vec{b} &\rfeq \left[ \, \ul{ b} \, , \, - \frac{\gpm}{M_A} P^+ b^-
    \, \right],
\end{align}
where $M_A \approx A m_N$.  Using the dictionary \eqref{e:NRvars}, we
can impose rotational invariance in the rest frame through $\vec{ \bar
  p} , \vec{b}$ to constrain the form of the Wigner distribution.
Then we can change variables back to the boost-invariant quantities
$\bar{\alpha} , P^+ b^-$ which describe the high-energy collision.

%###################################################################################

%\bibliographystyle{JHEP} 
%\bibliography{references}

\providecommand{\href}[2]{#2}\begingroup\raggedright\endgroup

\end{document}